\documentclass[11pt]{article}

\usepackage[final]{acl}

\usepackage[table]{xcolor} 
\newcommand{\up}[1]{\textcolor{teal}{\scriptsize~$\uparrow$#1}} 

\usepackage{algorithm}
\usepackage{algpseudocode}

\usepackage[many]{tcolorbox}
\usepackage{enumitem}

\newtcolorbox{PromptCard}[1][]{
    enhanced,
    title=Observer Prompt: Abridged Template,
    fonttitle=\bfseries\small, 
    fontupper=\small,          
    colframe=black!75,         
    colback=gray!5,            
    coltitle=white,
    boxrule=0.8pt,             
    arc=3pt,                   
    left=4pt, right=4pt, top=4pt, bottom=4pt, 
    width=\linewidth,          
    #1
}

\usepackage{amsmath}
\usepackage{amssymb}
\usepackage{booktabs}
\usepackage{tabularx} 
\usepackage{multirow}

\usepackage{times}
\usepackage{latexsym}

\usepackage[T1]{fontenc}

\usepackage[utf8]{inputenc}

\usepackage{microtype}

\usepackage{inconsolata}

\usepackage{graphicx}

\title{REAT: A Reflective Experience-Augmented Tutoring Framework for Multi-turn Mathematical Instruction}

\author{\textbf{Jianheng Zhou\textsuperscript{1}, Chaoli Zhang\textsuperscript{2}\thanks{Corresponding authors: Chaoli Zhang and Haoyang Li.}, Xingjun Wei\textsuperscript{3}, Xinliang Zhou\textsuperscript{4}, Giancarlo Fortino\textsuperscript{5},}\\
  \textbf{Xing Fan\textsuperscript{6}, Yanfeng Wang\textsuperscript{7}, Qingsong Wen\textsuperscript{8}, and Haoyang Li\textsuperscript{8}\textsuperscript{*}\thanks{This work was supported in part by the NSFC under Grant No. 62502456}}\\
  \normalfont
  \textsuperscript{1}School of Electrical and Electronic Engineering, Nanyang Technological University\\
  \textsuperscript{2}Zhejiang Normal University \quad
  \textsuperscript{3}Telfer School of Management, University of Ottawa\\
  \textsuperscript{4}Nanyang Technological University \quad
  \textsuperscript{5}University of Calabria\\
  \textsuperscript{6}APSS, Hong Kong Polytechnic University \quad
  \textsuperscript{7}School of Artificial Intelligence, Shanghai Jiao Tong University\\
  \textsuperscript{8}Squirrel Ai Learning\\
  \textbf{Correspondence:} \href{mailto:chaolizcl@zjnu.edu.cn}{chaolizcl@zjnu.edu.cn}, \href{mailto:derekli@squirrelai.com}{derekli@squirrelai.com}}

\begin{document}
\maketitle
\begin{abstract}
Current Large Language Models (LLMs) excel at solving complex mathematical problems, yet this proficiency does not inherently translate into effective tutoring. While advanced LLM tutors may leverage multi-agent frameworks or fine-tuning, most still lack a mechanism to systematically accumulate and reuse pedagogical experience over time, limiting their adaptability to diverse student needs during fluid, multi-turn interactions. To bridge this gap, we propose the Reflective Experience-Augmented Tutoring (REAT) framework, which couples experience distillation from historical dialogues with real-time adaptive retrieval. Driven by a multi-agent Observer-Critic-Mentor (OCM) distillation pipeline, REAT reviews past conversational trajectories and distills raw interactions into structured, problem-agnostic pedagogical experiences. During live tutoring, a state-aware retrieval module injects these curated experiences to provide adaptive scaffolding based on the student's cognitive state. Experiments demonstrate that the proposed framework significantly outperforms both prompt-only and supervised fine-tuning (SFT) baselines, particularly in improving complex, low-scoring tutoring scenarios. Crucially, the distilled experiences exhibit robust generalization across diverse model architectures and mathematical datasets.
\end{abstract}


\section{Introduction}

The rapid evolution of Large Language Models (LLMs) is fundamentally reshaping educational interactions \cite{chu-etal-2025-llm}. State-of-the-art models can achieve near-perfect accuracy on K-12 mathematics problems now, meaning mathematical problem-solving is no longer the main technological hurdle \cite{rele2023benchmark}. However, providing a correct answer does not guarantee effective tutoring. \cite{macina-etal-2025-mathtutorbench,maurya-etal-2025-unifying}. In multi-turn tutoring, an effective model requires pedagogical capabilities: continuous \textbf{scaffolding} to guide thinking \cite{vandepol2010scaffolding}, precise \textbf{attribution} of cognitive gaps, \textbf{empathy} to sustain student engagement, and dynamic \textbf{strategy adaptation}. 

\begin{figure*}[t]
\centering
  \includegraphics[width=0.85\textwidth]{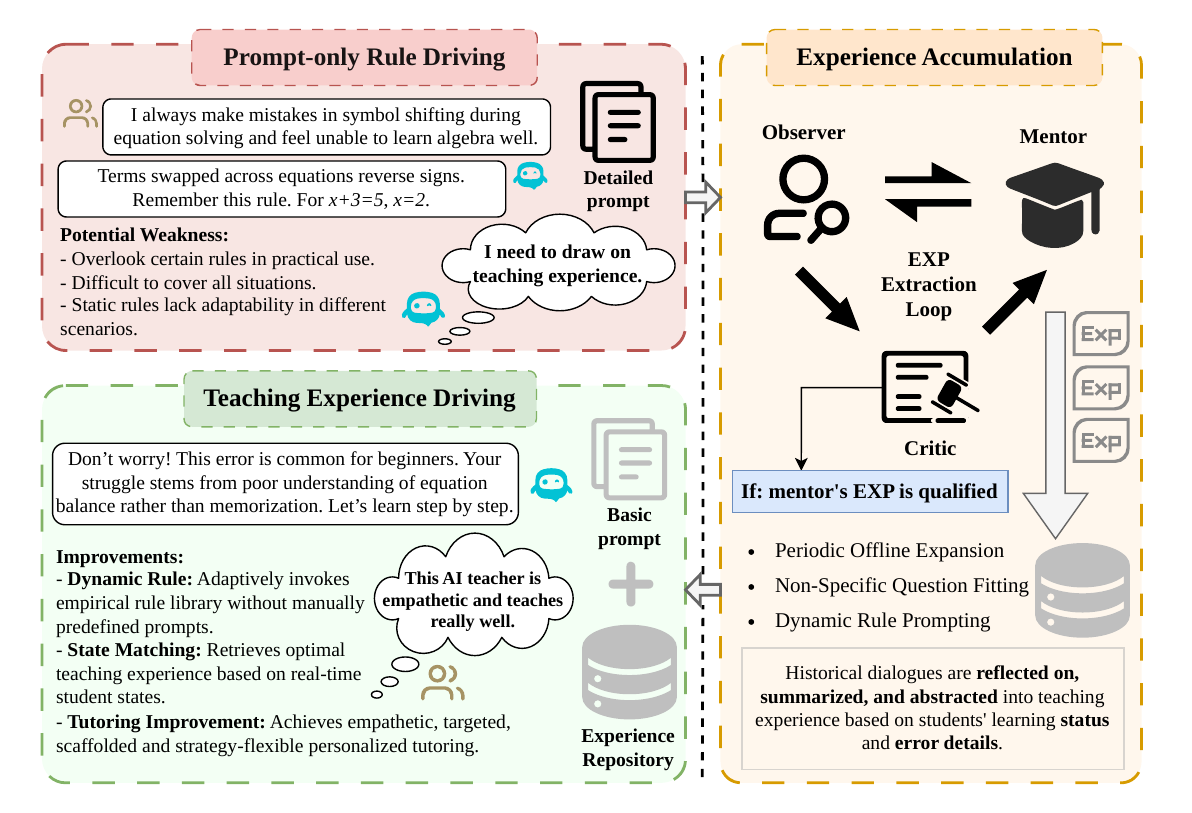}
  \caption{\textbf{Conceptual comparison between conventional static tutoring systems and our proposed REAT framework.} Unlike existing tutors bound by predefined, static mechanisms, our framework systematically accumulates and reuses pedagogical experiences over time to achieve dynamic and adaptive multi-turn tutoring.}
  \label{fig:intro}
\end{figure*}

Existing LLM tutors primarily rely on predefined prompts or vanilla retrieval, which often leads to relatively static strategies \cite{rw08,rw11,wang-etal-2024-problem}. While such rules enable basic socratic interactions, this reliance on static guidance makes it challenging to address unpredictable student confusion. Moreover, these static systems lack a continuous evolution mechanism. While recent work on agent reflection and memory shows that LLMs benefit from accumulating past feedback across trajectories \cite{rw18,rw23,rw24,rw25}, most current tutoring systems still fail to convert past interactions into reusable external memory. 

In practice, teachers often rely on expert mentors to reflect on and accumulate tutoring practices \cite{Larrivee01102000,schon2017reflective}. Inspired by this human reflection process, we propose a Reflective Experience-Augmented Tutoring (REAT) framework driven by a multi-agent Observer-Critic-Mentor (OCM) distillation pipeline. REAT systematically evaluates historical tutoring trajectories and abstracts them into structured pedagogical experiences. A rigorous gating mechanism admits only verified, high-quality experiences into the repository. During  tutoring, state-aware retrieval equips the teacher model with targeted guidance, enabling flexible strategy adjustments based on the student's cognitive state.

The main contributions of this work are summarized as follows:

    
    

\begin{itemize}
    \item \textbf{We propose the REAT framework.} Driven by a multi-agent OCM distillation pipeline, it evaluates historical trajectories and repairs weak responses, distilling raw interactions into reusable pedagogical experiences.
    
    \item \textbf{We construct a self-optimizing experience repository with state-aware retrieval.} The repository dynamically evolves through continuous deduplication, merging, and pruning. During tutoring, it retrieves state-aligned experiences to provide real-time and flexible scaffolding.
    
    \item \textbf{We validate the effectiveness and generalization of non-parametric pedagogical enhancement.} Experiments demonstrate that our approach significantly outperforms prompt-only and Supervised Fine-Tuning (SFT) baselines, particularly in improving challenging, low-scoring interactions. Crucially, the distilled experiences exhibit robust generalization across diverse model architectures and mathematical datasets.
\end{itemize}

\section{Related Work}
\label{sec:related_work}

\subsection{Intelligent Tutoring and Educational Dialogue Systems}

Before the era of Large Language Models (LLMs), early tutoring systems like CIRCSIM-Tutor, ITSPOKE, and Beetle II had already demonstrated that natural language dialogue could effectively diagnose student misconceptions and provide multi-turn feedback \cite{rw01,rw02,rw03}. AutoTutor further proved that conversational interaction leads to measurable learning gains, emphasizing the importance of adaptive instructional strategies and dialogue management \cite{rw04}. Later, with the shift toward data-driven modeling, benchmarks like CIMA \cite{rw05} and MathDial \cite{rw06} introduced rich pedagogical annotations to educational dialogues. These resources allowed researchers to study tutoring as a structured, step-by-step instructional process rather than just a simple response generation task.

\subsection{LLM-based Tutors and Evaluation Paradigms}
Recent research explores LLMs as conversational tutors, showing that strong problem-solving skills do not automatically guarantee effective pedagogy \cite{rw07}. To improve tutoring quality, current systems rely on personalized steering \cite{rw08} and structured hint generation \cite{rw09,rw10}. Concurrently, evaluation and interaction paradigms have shifted toward multi-agent architectures and user simulations \cite{rw11,rw12}. Decomposing the system into separate teacher, learner, and evaluator agents makes it much easier to model multi-round interactions and evaluate the tutoring process turn by turn \cite{rw12,rw14}. However, these multi-agent frameworks mostly focus on offline simulation or training optimization. They rarely use the reflective feedback generated during these interactions to guide teachers in real-time during  tutoring.

\subsection{Retrieval Augmentation, Reflection, and External Experience}

Recent advancements in language agents have increasingly leveraged retrieval-augmented generation (RAG), reflective refinement, and external memory. Standard RAG typically combines parametric models with external factual knowledge \cite{rw15}. In dialogue systems, incorporating external memory structures has proven effective for enhancing response informativeness \cite{rw16,rw17}. Furthermore, frameworks like Self-Refine and Reflexion show that LLMs can iteratively optimize their outputs using feedback and reflective memory \cite{rw18,rw19}. In the broader agentic domain, long-term memory and externalized experience accumulation have also become critical for sustained evolution \cite{rw20,rw21,rw22,rw23,rw24,rw25}. Unlike these general approaches, our framework focuses exclusively on pedagogical experience. Rather than retrieving factual knowledge or verbatim dialogue histories, our system distills high-quality instructional behaviors through offline reflection.

\section{Methodology}

\begin{figure*}[t]
\centering
  \makebox[\textwidth][c]{\includegraphics[width=1.1\textwidth]{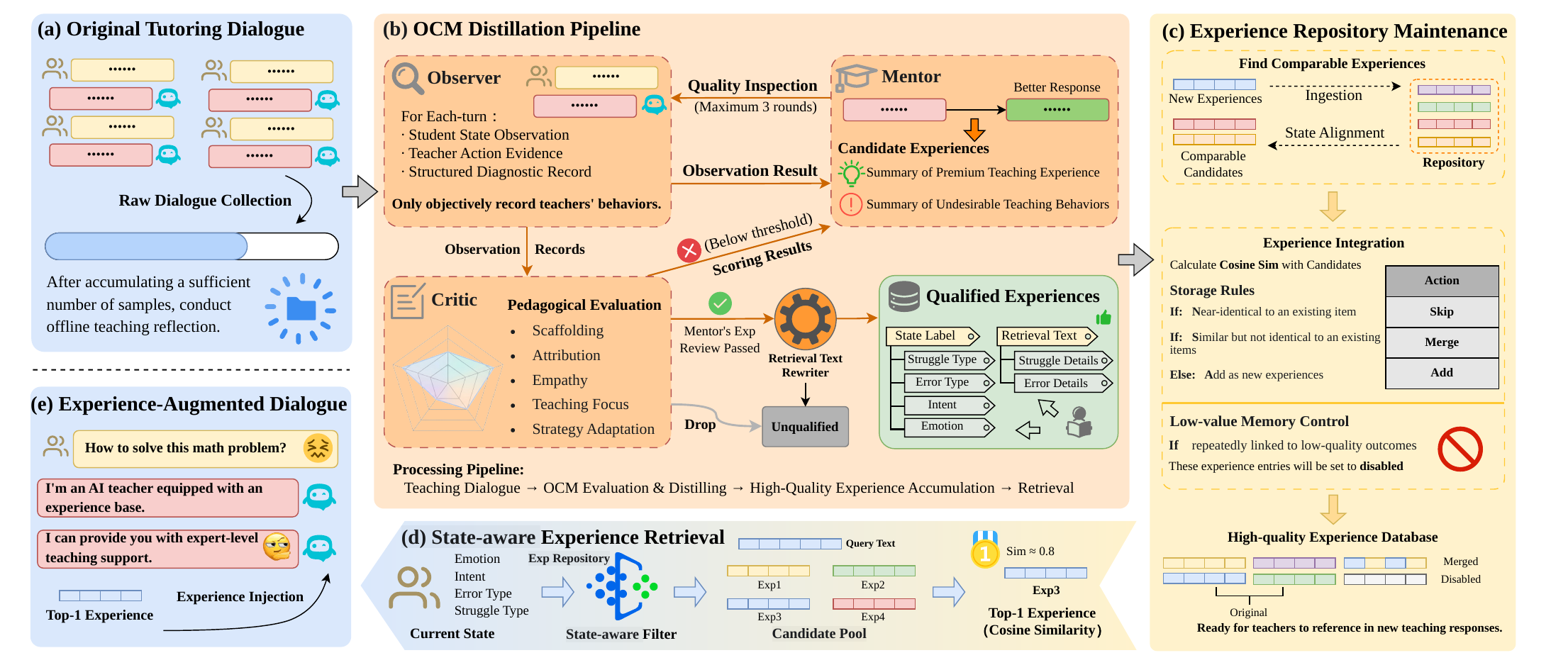}}
  \caption{\textbf{The overall architecture of the proposed REAT framework.} The figure illustrates the continuous life-cycle and flow of pedagogical experiences. Historical trajectories accumulated from (a) Original Tutoring Dialogue undergo rigorous  review and repair via (b)  OCM Distillation Pipeline. The abstracted high-quality experiences are then systematically organized and updated within an experience repository governed by (c). During  tutoring, the system utilizes (d) State-aware Experience Retrieval to dynamically match relevant experiences based on the student's current cognitive state. Finally, the retrieved experience is injected into (e) Experience-Augmented Dialogue, equipping the  AI Tutor with targeted and flexible pedagogical instruction.}
\label{fig:arch}
\end{figure*}

In this section, we formally define the multi-turn mathematics tutoring task. Let $P$ denote a given mathematics problem, and $H = \{(u_1, r_1), \dots, (u_{t-1}, r_{t-1})\}$ represent the dialogue history, where $u_i$ and $r_i$ are the student's utterance and the teacher's response at turn $i$, respectively. At turn $t$, our objective is to develop an automated tutoring system $f(\cdot)$ that generates the next teacher response $r_t$ based on $P$, $H$, and the current student utterance $u_t$:

\begin{equation}
r_t = f(P, H, u_t).
\end{equation}

Unlike traditional question answering, this task is not just about giving the correct answers. The main goal is to provide continuous tutoring support across multi-turn conversations. By tracking the student's understanding and adapting its guidance, the system helps them think independently and truly master the concepts.




\subsection{Overall Framework}
\label{sec:overall_framework}

To dynamically enhance the pedagogical reasoning of LLMs during multi-turn tutoring, we propose the \textbf{REAT framework} (Figure~\ref{fig:arch}). The architecture consists of two tightly coupled components: an execution module (the AI Tutor) that manages real-time interactions, and a multi-agent \textbf{OCM distillation pipeline} that drives pedagogical evolution by distilling high-quality instructional experiences from historical trajectories. To align the presentation with Figure~\ref{fig:arch}, the workflow is organized as: (a) original tutoring dialogue, (b) OCM distillation pipeline, (c) experience repository maintenance, (d) state-aware experience retrieval, and (e) experience-augmented dialogue.

During  tutoring, the base AI Tutor generates the response $r_t$ under a generative policy $\pi_{\theta}$. To move beyond static rules, this generation process is conditioned on $e_t$, a targeted pedagogical experience dynamically retrieved from the experience repository:
\begin{equation}
    r_t \sim \pi_{\theta}(\cdot \mid P, H_{<t}, u_t, e_t).
\end{equation}
The  OCM distillation pipeline continuously updates and controls the quality of the repository, ensuring the  tutor is constantly augmented by validated pedagogical priors.

\subsection{OCM Distillation Pipeline}
\label{sec:ocm_distillation}

The core of our REAT framework lies in the  OCM distillation pipeline, which simulates a rigorous, meta-cognitive pedagogical review. It systematically inspects past trajectories, executes targeted revisions, and distills validated behaviors into reusable experiences.

\paragraph{Observer.} 
The Observer acts as a diagnostic encoder. Rather than making a subjective quality judgment, it first maps the raw trajectory to a structured cognitive and behavioral state representation $s_t$:
\begin{equation}
s_t = \mathcal{O}(P, H_{<t}, u_t, r_t),
\end{equation}
where $\mathcal{O}(\cdot)$ denotes the mapping function. This state captures critical interaction evidence, including the types of student misconceptions and the corresponding teacher intervention.

\paragraph{Critic.} 
The Critic evaluates the pedagogical efficacy of the response $r_t$ grounded in $s_t$. We formalize this evaluation across five dimensions: scaffolding ($v_{scaf}$), attribution ($v_{attr}$), empathy ($v_{emp}$), teaching focus ($v_{foc}$), and strategy adaptation ($v_{ada}$). The Critic outputs a dimension score vector $\mathbf{v}_t$ alongside an aggregated pedagogical score $\mathcal{S}_t$:
\begin{equation}
\mathbf{v}_t, \mathcal{S}_t = \mathcal{C}(s_t, r_t).
\end{equation}
Conditioning on $s_t$ enables the Critic to accurately identify instructional flaws.

\begin{table}[t]
\centering
\small
\setlength{\tabcolsep}{3pt}
\caption{Operational definitions of the five pedagogical criteria used by the Critic.}
\label{tab:pedagogical_criteria}
\begin{tabular}{lp{0.62\columnwidth}}
\toprule
\textbf{Criterion} & \textbf{What it evaluates} \\
\midrule
Scaffolding & Preserve the student's cognitive agency through stepwise guidance rather than answer giving. \\
Attribution & Identify the student's actual misconception or reasoning bottleneck. \\
Empathy & Respond to affective signals in a way that sustains productive engagement. \\
Teaching focus & Stay aligned with the current learning obstacle and advance the dialogue. \\
Strategy adaptation & Adjust the intervention as evidence of struggle or progress accumulates. \\
\bottomrule
\end{tabular}
\end{table}

For clarity, the Observer represents each turn using practical teaching signals---student emotion, intent, error type, struggle status, and free-text diagnostics---rather than fine-grained psychological labels. Table~\ref{tab:pedagogical_criteria} makes explicit the pedagogical criteria used by the Critic; full score-level descriptions and the complete state schema are provided in Appendix~\ref{app:critic_rubric} and Appendix~\ref{app:student_state_schema}.

\paragraph{Rule-Based Routing.} 
To decouple evaluation logic from control flow, the decision to revise is governed deterministically. Given a global revision threshold $\tau_{rev}$ and a minimum dimension bound $\tau_{min}$, the routing indicator $D_t \in \{0, 1\}$ is defined as:
\begin{equation}
D_t = \mathbb{I}(\mathcal{S}_t < \tau_{rev} \lor \min(\mathbf{v}_t) < \tau_{min}).
\end{equation}
Trajectories triggering $D_t = 1$ exhibit pedagogical flaws and are routed for revision; otherwise, they proceed directly to distillation.

\paragraph{Mentor and Problem-Agnostic Distillation.} 
When activated, the Mentor uses the dialogue context and the diagnostic vector $\mathbf{v}_t$ to construct an improved response $r_t^*$:
\begin{equation}
r_t^* = \mathcal{M}(P, H_{<t}, u_t, s_t, \mathbf{v}_t).
\end{equation}
To prevent erroneous revisions from corrupting the experience repository $\mathcal{E}$, $r_t^*$ undergoes a secondary Critic inspection, yielding a new score $\mathcal{S}_t^*$. If it passes a strict quality gate $\gamma$, a distillation function $\mathcal{D}(\cdot)$ extracts the structured experience $e_{new}$. Crucially, to ensure that the experience can generalize to different mathematical problems, $\mathcal{D}(\cdot)$ rewrites the problem-specific diagnosis into a problem-agnostic description, removing explicit numbers and variables to abstract the core cognitive struggle:
\begin{equation}
e_{new} = \mathcal{D}(r_t^*),
\end{equation}
\begin{equation}
\mathcal{E} \leftarrow \mathcal{E} \cup \{e_{new}\} \quad \text{if } \mathcal{S}_t^* \ge \gamma.
\end{equation}

\subsection{Experience Representation and Retrieval}
\label{sec:experience_retrieval}

To transform the raw corpus into a generalizable experience repository, each validated experience is structured to capture its underlying tutoring logic, independent of the original problem formulation.

\paragraph{Experience Representation.} 
Formally, an experience $e \in \mathcal{E}$ is represented as a tuple:
\begin{equation}
e = \langle \mathbf{k}_e, \mathbf{v}_e \rangle, \quad \text{where } \mathbf{k}_e = \langle c_e, q_e \rangle.
\end{equation}
Here, the value $\mathbf{v}_e$ contains the abstracted teaching summary (e.g., the triggering difficulty, applied strategy, and scaffolding steps). The retrieval key $\mathbf{k}_e$ combines coarse-grained tags $c_e$ (e.g., error types, student emotion) and a dense semantic embedding $q_e$. Notably, $q_e$ is encoded directly from the problem-agnostic description of the student's core difficulty.

\paragraph{Coarse-to-Fine Retrieval.} 
During  tutoring, the system extracts the current state to form a target key $\mathbf{k}_t = \langle c_t, q_t \rangle$. The retrieval process has two stages. First, state-aware filtering selects candidates that match the coarse tags:
\begin{equation}
\mathcal{E}_{cand} = \{e \in \mathcal{E} \mid \text{Match}(c_t, c_e) = 1\}.
\end{equation}
Next, cosine similarity on the dense embeddings finds the most relevant teaching experience:
\begin{equation}
e_t = \arg\max_{e \in \mathcal{E}_{cand}} \cos(q_t, q_e).
\end{equation}
This two-stage method ensures retrieval prioritizes deep pedagogical alignment over superficial lexical matching. Rather than acting as a rigid template, the retrieved experience $e_t$ provides high-level guidance. By integrating this guidance with the dialogue context, the AI tutor dynamically adapts to the student's current state using verified instructional behaviors.

\subsection{Experience Repository Maintenance}
\label{sec:experience_maintenance}

To sustain retrieval efficiency and strategic diversity, the repository is continuously self-optimized via deduplication, consolidation, and utility pruning. Let $\text{Sim}(\cdot, \cdot)$ denote the semantic similarity function.

\paragraph{Deduplication and Merging.} 
Before integration, a candidate $e_{new}$ is compared against its corresponding state cluster. To prevent saturation with highly similar patterns, candidates exceeding a redundancy threshold $\tau_{dup}$ are directly discarded:
\begin{equation}
\text{Discard } e_{new} \quad \text{if } \exists e \in \mathcal{E}, \text{Sim}(e_{new}, e) \ge \tau_{dup}.
\end{equation}
Conversely, if a candidate shows conceptual overlap but contains complementary details, the system invokes a language model to synthesize a generalized experience, replacing the original entry:
\begin{equation}
\begin{split}
e_{merged} &= \text{Merge}(e_{new}, e) \\
&\quad \text{for } \tau_{merge} \le \text{Sim}(e_{new}, e) < \tau_{dup}.
\end{split}
\end{equation}

\paragraph{Utility Pruning.} 
The framework continually monitors  performance. Each experience maintains a utility score $U(e)$. When an experience is retrieved at turn $k$, its utility is updated using the subsequent  evaluation score $\mathcal{S}_{eval}^{(k)}$ via an exponential moving average:
\begin{equation}
U^{(k)}(e) = \alpha U^{(k-1)}(e) + (1 - \alpha) \mathcal{S}_{eval}^{(k)}.
\end{equation}
Experiences that consistently yield sub-optimal outcomes decay in utility. Those falling below a threshold $\tau_{prune}$ are removed:
\begin{equation}
\mathcal{E} \leftarrow \mathcal{E} \setminus \{e \mid U^{(k)}(e) < \tau_{prune}\}.
\end{equation}
These operations systematically prioritize teaching quality and long-term utility over pure data accumulation.

\section{Experiments}

\subsection{Experimental Setup}
\label{sec:experimental_setup}

\subsubsection{Model Configurations}
\label{sec:model_configurations}

To evaluate our framework, we primarily employ the proprietary \textbf{Doubao-seed} model as our main backbone for core system validation and iterative experience repository construction. To demonstrate the broad applicability of the OCM distillation pipeline across different architectures, we also include the open-source \textbf{Qwen3-8B} \cite{qwen3} and an additional strong proprietary model, \textbf{DeepSeek-v3} \cite{deepseekv3}, in our controlled comparisons under prompt-only and experience-augmented settings. Additionally, we include a Supervised Fine-Tuning (SFT) version of Qwen3-8B to contrast parametric fine-tuning with our non-parametric approach. GPT-5 drives the  multi-agent OCM distillation pipeline, while GPT-5-mini handles problem-agnostic rewriting and merging.

\subsubsection{Datasets and Experience Construction}
\label{sec:datasets}

We primarily use \textbf{GSM8K} \cite{gsm8k} for multi-turn mathematical tutoring experiments. To test generalization, we also evaluate on \textbf{APE210K} \cite{ape210k} and \textbf{Math23K} \cite{math23k}. 

We build the experience repository iteratively using the first 500 problems from the \textbf{GSM8K training set}. We process these in five sequential batches of 100 problems. Each batch acts as an unseen test set for the previous one. For final testing, we select three independent subsets (100 problems each) from the \textbf{GSM8K test set}.

\subsubsection{Evaluation and Simulation Protocol}
\label{sec:evaluation_protocol}

Since existing conversational benchmarks are not well-suited for multi-turn mathematical tutoring, we employ an internal OCM pipeline-based evaluation system tailored to our pedagogical objectives (rubric in Appendix \ref{app:critic_rubric}). Furthermore, we conduct cross-architecture validation and human-assisted verification to mitigate model-specific bias. For realistic interactions, we utilize DeepSeek-v3 to generate diverse student personas, which are kept strictly identical for a given problem across all baselines to ensure fair comparison (Appendix \ref{sec:implementation_details}).

\paragraph{Controlled Comparison Protocol.}
For every evaluated problem, all systems interact with the same preassigned student persona and use the same teacher backbone and base prompt. The only treatment difference is whether the retrieved experience fields are populated, isolating the contribution of experience augmentation from prompt wording and simulation variability. We construct the repository from the first 500 GSM8K training problems in five sequential batches of 100 problems, and report final results on three independent 100-problem subsets of the GSM8K test set.

\begin{table*}[htbp]
\centering
\small
\setlength{\tabcolsep}{12pt} 
\caption{Turn-level evaluation results of different teacher models on the GSM8K test splits. We compare the prompt-only baselines, supervised fine-tuning (SFT), and our proposed REAT framework across various model backbones. Green arrows ($\uparrow$) indicate absolute improvements over the respective prompt-only baseline.}
\label{tab:main_results}
\begin{tabular}{lcccc}
\toprule
\textbf{Teacher Model} & \textbf{Test-1} & \textbf{Test-2} & \textbf{Test-3} & \textbf{Avg.} \\
\midrule
Prompt-only Qwen3-8B      & 80.49 & 81.30 & 80.06 & 80.62 \\
SFT Qwen3-8B              & 80.78 & 81.19 & 80.32 & 80.76 \\
\rowcolor{gray!10} 
REAT Qwen3-8B             & 82.26 \up{1.77} & 82.95 \up{1.65} & 83.04 \up{2.98} & 82.75 \up{2.13} \\
\midrule
Prompt-only DeepSeek      & 89.60 & 90.82 & 88.74 & 89.72 \\
\rowcolor{gray!10} 
REAT DeepSeek             & 92.59 \up{2.99} & 94.46 \up{3.64} & 92.71 \up{3.97} & 93.25 \up{3.53} \\
\midrule
Prompt-only Doubao        & 88.28 & 91.18 & 88.31 & 89.26 \\
\rowcolor{gray!10} 
\textbf{REAT Doubao}             & 92.98 \up{4.70} & 94.03 \up{2.85} & 93.53 \up{5.22} & 93.51 \up{4.25} \\
\bottomrule
\end{tabular}
\end{table*}

\subsection{Main Results}
\label{sec:main_results}

To evaluate our framework, we instantiate it across various LLM backbones. In the following experiments, we denote these experience-augmented models as \textit{REAT model} (e.g., \textit{REAT Doubao}), distinguishing them from the \textit{Prompt-only} and \textit{SFT} baselines.

\begin{figure}[htbp]
\centering
  \includegraphics[width=1.0\columnwidth]{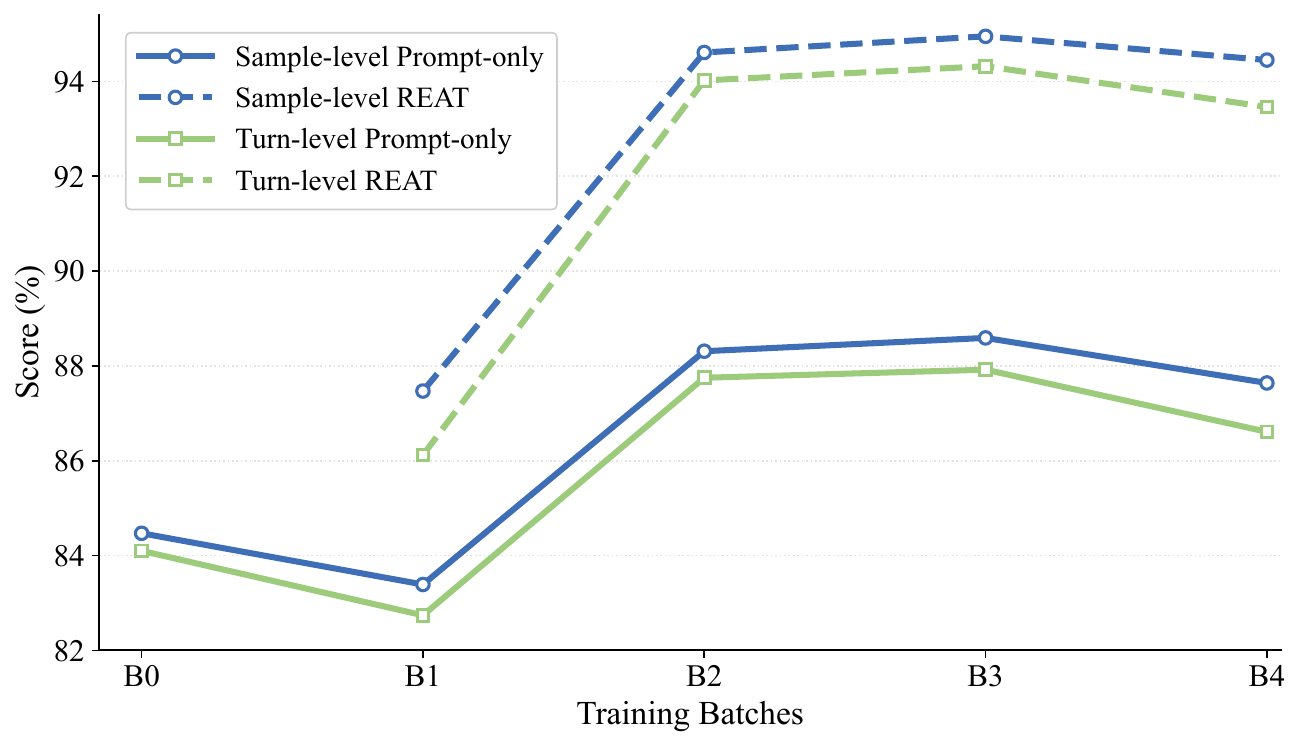}
  \caption{Performance comparison across training batches. Note: REAT values are absent at B0 due to the cold-start phase lacking an established experience repository.}
  \label{fig:train_batches_levels}
\end{figure}

We first examine the trajectory of experience accumulation during the training phase (Figure~\ref{fig:train_batches_levels}). Evaluation performance is measured at two levels: the sample level (evaluating the complete instructional dialogue for an entire problem) and the turn level (evaluating individual instructional turns). At the cold-start stage (B0), the baseline system achieves 84.47\% at the sample level and 84.10\% at the turn level. As the experience repository grows (B1 to B4), the experience-augmented teacher (REAT) consistently outperforms the prompt-only baseline. By the second batch (B2–B4), performance plateaus, consistently outperforming the baseline by over 6.2 points. This indicates that the proposed framework successfully constructs a reusable repository, providing sustained tutoring benefits beyond the initial phase.

Table~\ref{tab:main_results} presents results on three held-out test splits. Focusing primarily on the stronger proprietary backbone, \textbf{Doubao-seed}, the prompt-only baseline averages 89.26. Integrating our REAT framework boosts this average performance to 93.51. This absolute improvement of 4.25 points remains stable across all test subsets, demonstrating that explicitly distilling and reusing pedagogical experiences substantially enhances the instructional capabilities of already-capable LLMs.

\begin{figure*}[t]
    \centering
    \includegraphics[width=0.46\textwidth]{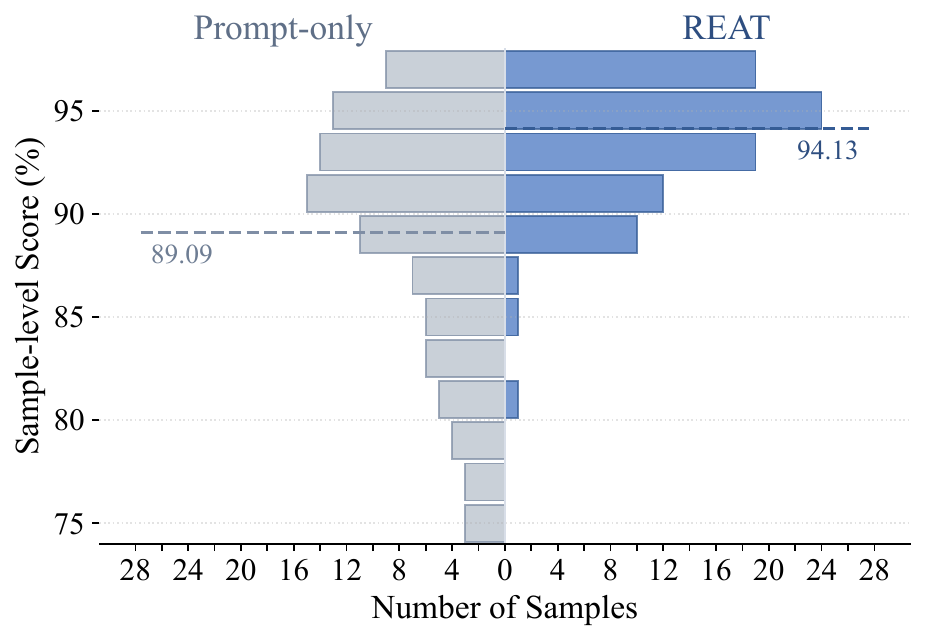}\hfill
    \includegraphics[width=0.46\textwidth]{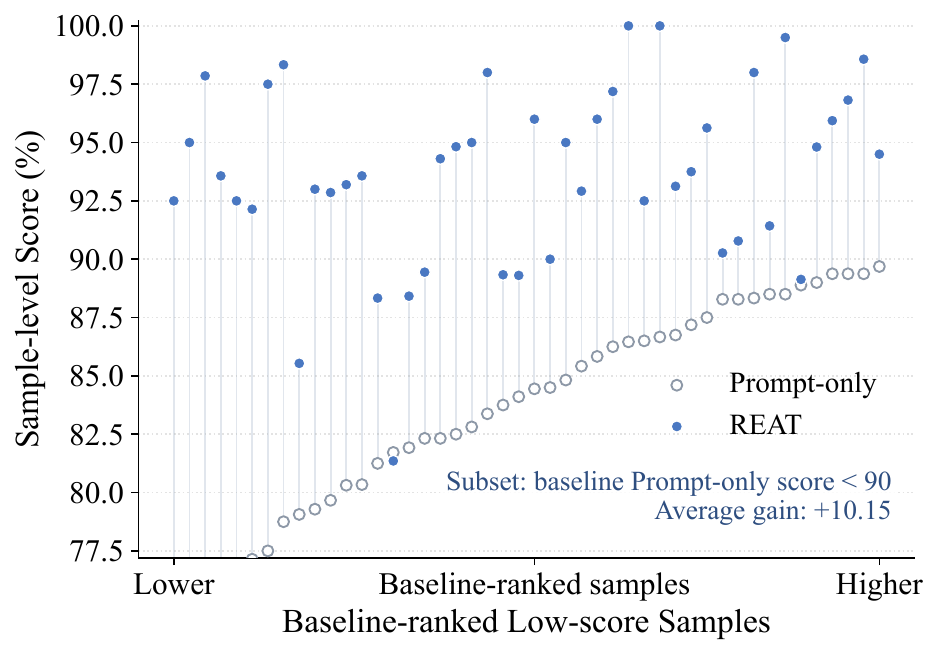}
    \caption{Where REAT improves tutoring quality on Test-3. \textbf{Left:} mirrored sample-level score distributions; dashed lines mark the respective averages. \textbf{Right:} paired scores for challenging cases with prompt-only score below 90; each line connects the same sample before and after augmentation.}
    \label{fig:distribution_shift_rescue}
\end{figure*}

\textbf{Distribution Shift and Low-Score Rescue.} Figure~\ref{fig:distribution_shift_rescue} shows that REAT shifts the overall sample-level score distribution upward, raising the Test-3 average from 89.09 to 94.13 and substantially reducing the low-score tail. The effect is particularly strong for challenging cases: among samples whose prompt-only score is below 90, REAT yields an average gain of 10.15 points. This analysis shows that the aggregate improvement is not driven only by already successful interactions; retrieved experiences are especially useful when the base tutor struggles to provide adequate scaffolding. Additional paired case studies are presented in Appendix~\ref{app:case_studies}.

Besides, we evaluate the \textbf{Qwen3-8B} open-source model. While retrieval augmentation improves its prompt-only baseline (from 80.62 to 82.75), the relative gain is smaller compared to Doubao. We hypothesize that executing complex pedagogical strategies inherently requires stronger foundational language comprehension, aligning with recent findings on capacity bottlenecks \cite{rw25}. Additionally, SFT on Qwen3-8B yields negligible improvement (80.76), suggesting that non-parametric experience augmentation is a more effective adaptation route for pedagogical reasoning than parametric style transfer. Overall, empirical results confirm that reflective experience accumulation provides stable enhancements, particularly for high-capacity models capable of interpreting advanced instructional guidance.

\subsection{Judge Validation and Human Evaluation}
\label{sec:cross_judge_consistency}

Since the pedagogical experiences in our framework are distilled through the primary OCM pipeline, we need to verify that the observed gains are not biased by the evaluator. Therefore, we compare prompt-only and REAT across three distinct evaluation views: the primary GPT-5 OCM judge, an external Doubao OCM judge, and human expert annotators (guidelines in Appendix \ref{app:human_annotation}).

Figure~\ref{fig:cross_judge_trend} demonstrates a consistent improvement pattern across all views. Under the primary GPT-5 judge, REAT consistently outperforms prompt-only in every test split, yielding gains of 1.79 to 5.04 at the sample level and 2.85 to 5.22 at the turn level. The external Doubao judge corroborates this trend despite differing absolute scales, showing sample- and turn-level gains of 2.53 to 3.07 and 2.89 to 3.88 points, respectively.

Human evaluation closely aligns with these automated metrics. To ensure a rigorous yet cost-effective comparison, three experts evaluated a shared test subset (Test-1). All annotators consistently rated REAT higher, with the average turn-level score rising by 6.54 points (from 85.30 to 91.84), and the sample-level score by 5.94 points (from 86.56 to 92.50).

Beyond scalar scores, the annotators also conducted paired preference judgments and sanity checks for both automated scores and inferred learner states. These complementary checks help distinguish improvements in perceived pedagogical quality from artifacts caused by evaluator calibration or state labeling; their detailed outcomes are summarized in Figure~\ref{fig:cross_judge_trend} and Table~\ref{tab:human_sanity_checks}.
Pairwise preferences directly test which response experts would choose for the same learner context. The plausibility checks further assess whether the state signals that guide retrieval remain meaningful to independent experts.

\begin{figure}[H]
    \centering
    \includegraphics[width=\linewidth]{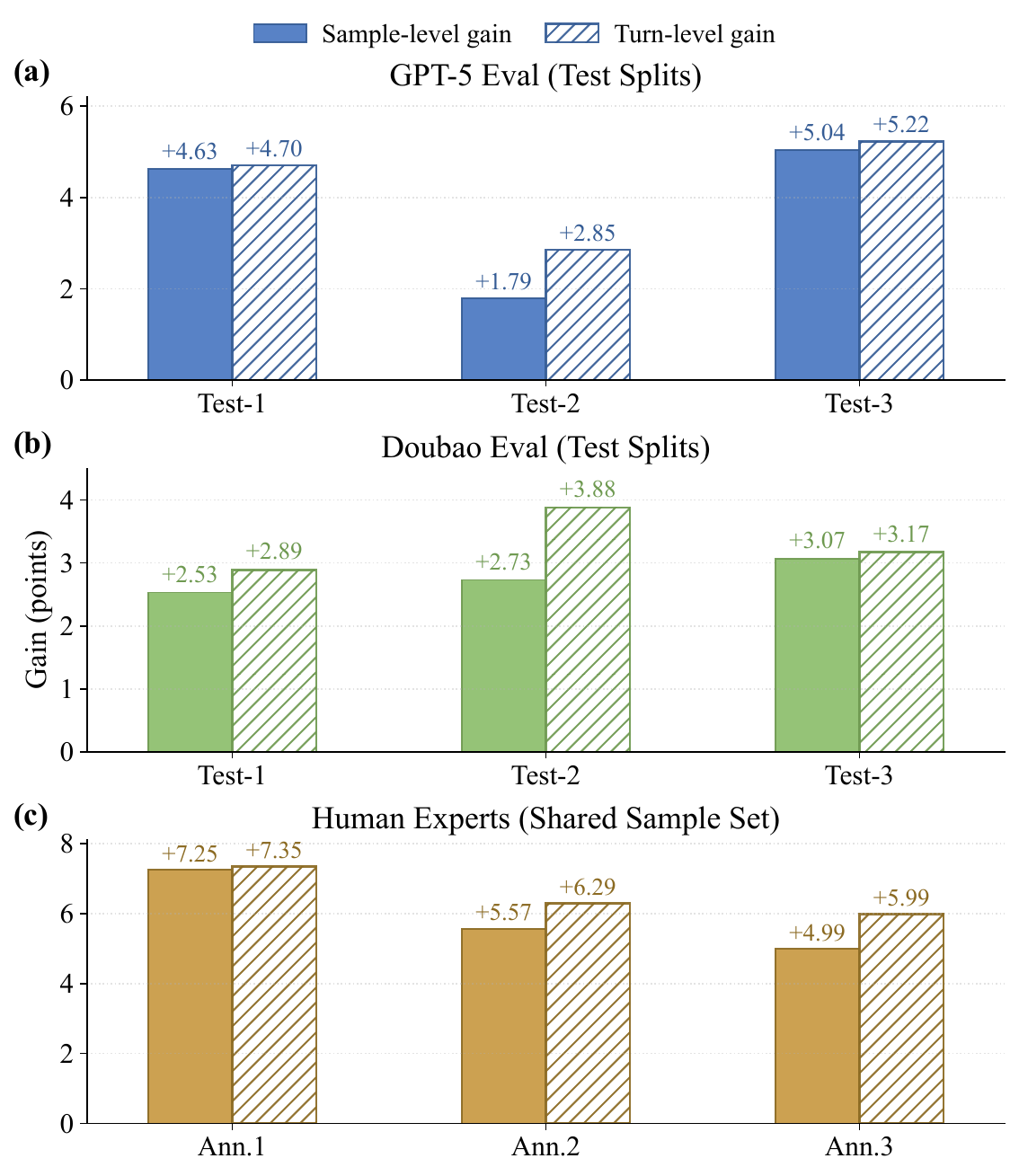}
    \caption{Trend consistency of relative gains achieved by REAT. \textbf{(a) and (b)} illustrate score improvements on the three test splits under the GPT-5 and Doubao judges. \textbf{(c)} shows gains assessed by three human experts on a shared subset. Absolute scores are detailed in Table~\ref{tab:cross_judge_absolute_scores}.}
    \label{fig:cross_judge_trend}
\end{figure}

\begin{table}[H]
\centering
\scriptsize
\setlength{\tabcolsep}{1.8pt}
\renewcommand{\arraystretch}{1.12}
\caption{Expert validation by annotator (\%). ``Gain'' is REAT minus prompt-only. Full statistics are in Appendix~\ref{app:human_eval_summary}.}
\label{tab:human_sanity_checks}
\begin{tabular}{p{0.36\columnwidth} c c c c}
\toprule
\textbf{Metric} & \textbf{Ann. 1} & \textbf{Ann. 2} & \textbf{Ann. 3} & \textbf{Avg.} \\
\midrule
Sample gain & 7.25 & 5.57 & 4.99 & 5.94 \\
Turn gain & 7.35 & 6.29 & 5.99 & 6.54 \\
\midrule
REAT preferred & 81.03 & 58.33 & 55.17 & 64.84 \\
Uncertain & 15.52 & 35.00 & 32.76 & 27.76 \\
Prompt-only preferred & 3.45 & 6.67 & 12.07 & 7.40 \\
\midrule
LLM score reasonable & 75.34 & 93.04 & 66.01 & 78.13 \\
State label plausible & 89.85 & 96.48 & 75.77 & 87.37 \\
\bottomrule
\end{tabular}
\end{table}

Figure~\ref{fig:cross_judge_trend} and Table~\ref{tab:human_sanity_checks} jointly test whether gains depend on the primary evaluator. Experts prefer REAT in 64.84\% of paired comparisons, versus 7.40\% for prompt-only, and judge the automated scores and inferred student states reasonable in 78.13\% and 87.37\% of cases, respectively. Together, these results confirm that the observed gains are robust across external judging and human assessment rather than relying on a single evaluator.

Importantly, we compare relative gains rather than raw scores because the GPT-5, Doubao, and human evaluators use different scoring ranges. Their agreement on the direction of improvement, together with the preference results, indicates that the benefit reflects perceived pedagogical quality rather than the calibration of any single judge.

\begin{figure*}[htbp]
\centering
  \includegraphics[width=0.96\textwidth]{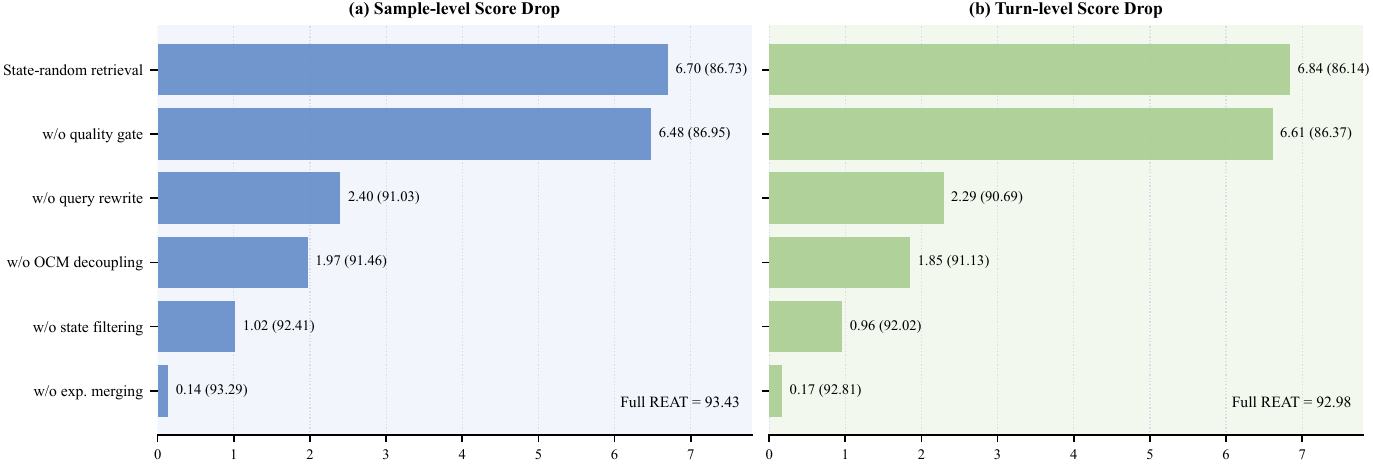}
  \caption{Ablation results for REAT. Each panel shows the absolute performance drop from full REAT. \textit{State-random} preserves state matching but randomly selects an experience from the matched candidates.}
  \label{fig:ablation}
\end{figure*}

\subsection{Ablation Study}
\label{sec:ablation}

To isolate the contribution of each core component, we evaluate several ablated variants of our framework. Figure~\ref{fig:ablation} shows the performance drop of each variant compared to the full REAT framework.

\textbf{Quality Gate.} Removing the strict quality gate causes the most severe performance drop ($-6.48$ sample-level, $-6.61$ turn-level). This confirms that  quality control is not just a maintenance step, but a critical safeguard against adding noisy or flawed teaching experiences into the repository.

\textbf{Problem-Agnostic Rewriting and OCM Decoupling.} Eliminating the rewriting step and merging the separated OCM roles into a single reflection process lead to notable sample-level drops of $2.40$ and $1.97$ points, respectively. This validates two key design choices: rewriting specific diagnoses into general retrieval cues improves downstream matching, and explicitly separating observation, scoring, and revision yields higher-quality distillation.

\textbf{State-Aware Filtering and Experience Merging.} Omitting coarse state-aware filtering in favor of pure semantic matching degrades sample-level performance by $1.02$, highlighting the necessity of anchoring retrieval in pedagogical alignment rather than superficial similarity. In contrast, removing semantic merging causes only a marginal drop ($-0.14$), aligning with its primary purpose of maintaining repository compactness and reducing redundancy rather than directly driving metric gains.

\textbf{State-Random Retrieval.} We further isolate online retrieval selection with \textit{State-random}, which preserves state matching but randomly selects an experience from the matched candidates. Figure~\ref{fig:ablation} shows that this variant drops to 86.73 at the sample level and 86.14 at the turn level---below the prompt-only baseline. Thus, injecting a high-quality but mismatched experience can be harmful. Together with the drops caused by removing query rewriting and the quality gate, this result shows that REAT's gains arise from the combination of quality-controlled experience construction and accurate state-aware selection, rather than from either component alone.

Overall, these results demonstrate that our framework's effectiveness does not come from retrieval alone. Instead, it emerges from the synergy of role-specialized reflection, rigorous quality gating, and state-aware experience augmentation.

\subsection{Generalization Analysis}
\label{sec:generalization}

To assess transfer across teacher backbones and unseen data, we use the repository built by the \textit{Doubao-seed} teacher (full plots in Appendix~\ref{app:additional_generalization}, Figure~\ref{fig:generalization}). It improves both sample- and turn-level scores on Claude and DeepSeek-v3 by roughly 2--4 points; DeepSeek-v3 is an independent teacher in this test despite serving as the student simulator during data generation. The same repository also improves both metrics by 3--4 points on APE210K and Math23K.

\textbf{OCM-Driver Robustness and Llama Transfer.} With the teacher backbone fixed, lightweight repositories distilled from the first two batches by Qwen-30B and DeepSeek-V3 improve over the 88.80/88.28 prompt-only baseline, reaching 89.92/89.43 and 91.06/91.45 sample/turn, respectively. The full GPT-5-driven repository reaches 93.43/92.98, indicating that OCM capability affects the quality ceiling but is not required for a benefit. Under the same lightweight protocol, REAT also improves Llama-3.1-70B from 82.83 to 84.41 at the sample level and from 82.49 to 83.92 at the turn level.

Together, these transfers show that REAT reuses pedagogical strategies across model families and data distributions instead of depending on one fixed teacher or dataset.




\section{Conclusion}
\label{sec:conclusion}


REAT couples OCM-based experience distillation with state-aware retrieval. It improves tutoring quality across models, datasets, and difficult interactions. By separating offline quality-controlled experience construction from online state-aware selection, REAT makes the source of each improvement explicit and reusable. The resulting design offers a scalable path from static prompting to adaptive tutors that accumulate verified pedagogical strategies.

\section*{Limitations}

Our evidence is based on simulated mathematical tutoring rather than real learner outcomes. Although external judges and human experts corroborate the quality gains, independent classroom studies remain necessary. We also evaluate only mathematics and short-to-medium tutoring horizons; transfer to open-ended subjects, multimodal settings, and long-term personalization remains to be established. Finally, the OCM pipeline and human evaluation incur practical cost, and repository behavior at production scale requires further study.

\paragraph{Potential Risks.}
Although our work focuses on mathematical tutoring, it may still introduce risks if deployed without oversight. While our state-aware module effectively captures general cognitive states, its granularity may be insufficient for students with complex psychological vulnerabilities or atypical emotional needs (e.g., depression or severe anxiety). In such edge cases, the retrieved pedagogical strategies might fail to provide the necessary specialized emotional support. Additionally, over-reliance on an AI tutor may weaken students’ independent problem-solving habits. These limitations suggest that such systems should be used with careful monitoring and should not replace human educators, especially when dealing with vulnerable student populations.

\section*{Use of AI Assistants}
AI assistants were used in a limited supporting role for coding assistance, translation to English, and checking grammar and logical flow. The final writing judgments were made and verified by the authors, who take full responsibility for the content of the paper.

\bibliography{custom}

\appendix

\section{Additional Generalization Results}
\label{app:additional_generalization}

\begin{figure}[htbp]
\centering
  \includegraphics[width=0.95\columnwidth]{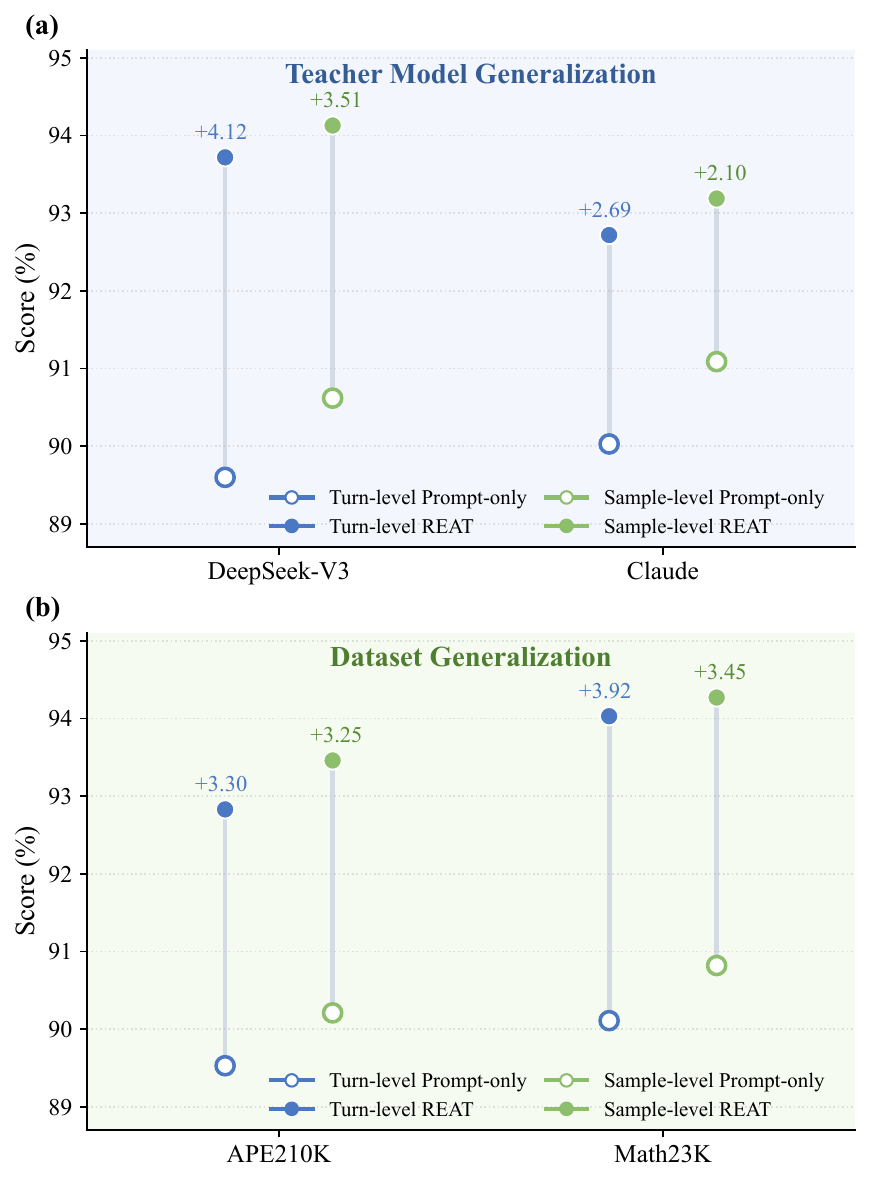}
  \caption{Generalization results of the proposed framework. \textbf{(a):} Cross-model generalization. \textbf{(b):} Cross-dataset generalization.}
  \label{fig:generalization}
\end{figure}

\section{Pedagogical Scoring Rubric}
\label{app:critic_rubric}

The Critic evaluates each tutoring turn along five pedagogically motivated dimensions: \textit{scaffolding}, \textit{attribution}, \textit{empathy}, \textit{teaching focus}, and \textit{strategy adaptation}. Each dimension is scored on a 0--4 scale, where higher scores indicate stronger pedagogical quality. The rubric is designed to distinguish genuine expert-like tutoring from superficially helpful but pedagogically weak responses.

\paragraph{Scaffolding.}
This dimension measures whether the teacher preserves the student's cognitive agency instead of completing the reasoning on the student's behalf.
\begin{itemize}[leftmargin=18pt, noitemsep, topsep=2pt]
    \item[\textbf{4:}] Strong stepwise guidance that helps the student make the key inference independently.
    \item[\textbf{3:}] Mostly appropriate guidance with minor over-explanation.
    \item[\textbf{2:}] Weak or pseudo-scaffolding, such as asking rhetorical questions and then immediately answering them, or heavily narrowing the reasoning path.
    \item[\textbf{1:}] Severe over-guidance that substantially reduces student thinking.
    \item[\textbf{0:}] Direct answer giving or full solution dumping.
\end{itemize}

\paragraph{Attribution.}
This dimension evaluates whether the teacher correctly identifies the student's actual misconception, reasoning gap, or source of confusion.
\begin{itemize}[leftmargin=18pt, noitemsep, topsep=2pt]
    \item[\textbf{4:}] Precise diagnosis of the student's true bottleneck.
    \item[\textbf{3:}] Mostly correct diagnosis with limited abstraction or specificity.
    \item[\textbf{2:}] Partially relevant but shallow diagnosis.
    \item[\textbf{1:}] Largely misaligned diagnosis.
    \item[\textbf{0:}] Complete failure to identify the relevant problem or diagnosis of the wrong issue.
\end{itemize}

\paragraph{Empathy.}
This dimension measures whether the teacher responds to the student's affective state in a genuine and instructionally useful way.
\begin{itemize}[leftmargin=18pt, noitemsep, topsep=2pt]
    \item[\textbf{4:}] Targeted emotional acknowledgment that supports continued engagement.
    \item[\textbf{3:}] Supportive but somewhat generic encouragement.
    \item[\textbf{2:}] Formulaic or weak empathy with limited pedagogical value.
    \item[\textbf{1:}] Perfunctory acknowledgment.
    \item[\textbf{0:}] Clear frustration or confusion is ignored altogether.
\end{itemize}

\paragraph{Teaching Focus.}
This dimension evaluates whether the teacher stays aligned with the student's current bottleneck and advances the dialogue in a focused manner.
\begin{itemize}[leftmargin=18pt, noitemsep, topsep=2pt]
    \item[\textbf{4:}] Remains tightly focused on the current learning obstacle.
    \item[\textbf{3:}] Mostly relevant guidance with minor drift.
    \item[\textbf{2:}] Unnecessary explanation or partial loss of focus.
    \item[\textbf{1:}] Substantial drift away from the student's actual need.
    \item[\textbf{0:}] Misses the student's problem entirely.
\end{itemize}

\paragraph{Strategy Adaptation.}
This dimension captures whether the teacher meaningfully adjusts instructional strategy when the student repeatedly struggles or fails to improve.
\begin{itemize}[leftmargin=18pt, noitemsep, topsep=2pt]
    \item[\textbf{4:}] Clear strategic adaptation (e.g., shifting from direct prompting to analogy or decomposition).
    \item[\textbf{3:}] Noticeable but incomplete adaptation.
    \item[\textbf{2:}] Superficial wording changes without a real pedagogical shift.
    \item[\textbf{1:}] Strong rigidity across turns.
    \item[\textbf{0:}] Complete failure to adapt despite repeated evidence that the previous strategy is not working.
\end{itemize}
\textit{Note: If the available history is insufficient to determine whether adaptation is needed, this dimension may be marked as not applicable.}

\paragraph{Operational Rules.}
Beyond scoring each dimension, the Critic follows specific operational rules. Major tutoring failures will automatically cap the score for that category. For instance, pseudo-scaffolding heavily penalizes the scaffolding score, fake empathy limits the empathy score, and misdiagnosing the student restricts the attribution score. The Critic also uses these scores to decide the next step: high-quality turns go straight to experience distillation, while weak or mediocre turns are sent to the Mentor for revision.

\paragraph{Purpose of the Rubric.}
The rubric is designed to reward good tutoring. Specifically, behaviors that encourage independent thinking, correctly identify errors, offer genuine emotional support, stay focused on the immediate problem, and adapt when a student keeps struggling. Ultimately, it serves as both an evaluation tool and a strict quality-control mechanism for building the experience repository in the OCM pipeline.

\section{Implementation Details}
\label{sec:implementation_details}

We use publicly available datasets, embedding models, and software libraries under their original research licenses or usage terms. Our experiments are conducted for research purposes only, and any future release of code or derived resources will need to remain consistent with the corresponding artifact licenses and access conditions.

\subsection{Model and System Configuration}
The  OCM pipeline (Observer, Critic, and Mentor) uses GPT-5 to provide detailed pedagogical diagnosis and quality-controlled experience distillation. To maintain evaluation consistency, we set the temperature to 0. For the  tutoring interaction, the teacher model is powered by Doubao-seed, while the student simulator uses DeepSeek-V3. To strictly isolate the impact of experience augmentation, the teacher prompt remains identical across both the prompt-only and REAT settings. They differ only in whether the retrieved experience fields contain data, which effectively prevents prompt wording from skewing the results.

Our external experience repository uses \textbf{FAISS} \cite{faiss} for efficient nearest-neighbor search, with semantic embeddings computed via \textbf{paraphrase-multilingual-MiniLM-L12-v2} \cite{sentence} for robust multilingual matching. Retrieval follows a coarse-to-fine approach: candidate experiences are first filtered by structured student-state attributes, and then ranked by cosine similarity. To query this index, a lightweight GPT-5 module rewrites the Observer's raw diagnostic fields into a generalized description of the student's cognitive bottleneck. This abstraction reduces reliance on the exact problem text, ensuring that retrieval is driven by true pedagogical similarity rather than mere lexical overlap.

\subsection{Online Latency Analysis}
\label{app:latency}

The OCM distillation pipeline runs entirely offline and therefore adds no deployment-time latency. Table~\ref{tab:latency} reports median online latency per tutoring round, excluding simulated-student generation. The main overhead comes from LLM/API calls for state diagnosis and query rewriting; embedding and FAISS search take approximately 0.03 seconds and are not the bottleneck. These measurements reflect a research prototype focused on tutoring quality; practical latency can be reduced by using a smaller Observer, merging diagnosis and rewriting calls, caching repeated states, streaming teacher output, or parallelizing independent steps.

\begin{table}[t]
\centering
\small
\setlength{\tabcolsep}{5pt}
\caption{Median online latency per tutoring round.}
\label{tab:latency}
\begin{tabular}{lc}
\toprule
\textbf{Component} & \textbf{p50 latency} \\
\midrule
Observer (state diagnosis) & 12.4s \\
Retrieval submodule & $\sim$4.2s \\
Teacher generation & 2.8s \\
Prompt assembly and orchestration & $\sim$0.8s \\
\midrule
\textbf{REAT total} & \textbf{$\sim$20.2s} \\
\bottomrule
\end{tabular}
\end{table}

\subsection{High-quality Dialogue Selection and LoRA Fine-tuning}
\label{app:lora_finetuning}

To provide a fine-tuned baseline for comparison against our experience-augmentation framework, we train the 8B teacher backbone using standard LoRA. Supervised training data is created by filtering the training split with our OCM pipeline, retaining only high-scoring teacher responses. Each training instance predicts the teacher's response based on the problem, dialogue history, and current student utterance via standard next-token prediction. This baseline operates without retrieval during inference, representing an alternative paradigm where high-quality pedagogical behaviors are embedded directly into model parameters rather than accessed from an external base.

\subsection{Student Persona Design}
\label{app:student_personas}

To simulate diverse tutoring trajectories, the student model is guided by a persona sampled from a predefined library. These personas act as practical dialogue settings rather than strict psychological profiles. Each specifies a forced first-turn trigger to consistently introduce a learning difficulty and a behavioral tendency for later turns. This ensures the system faces realistic tutoring challenges (e.g., misconception repair, emotional regulation) rather than a uniformly cooperative student.

The library includes nine representative student types:
\begin{itemize}[leftmargin=12pt, itemsep=2pt, topsep=2pt]
    \item \textbf{Passive Dependent}: Relies heavily on explicit step-by-step instructions; struggles with independent global planning.
    \item \textbf{Stubborn Impatient}: Insists on incorrect reasoning and becomes frustrated by repetitive or uninformative feedback.
    \item \textbf{Careless Reader}: Misreads or extracts incorrect numbers and conditions from the problem statement.
    \item \textbf{Blind Guesser}: Avoids genuine reasoning, offering unsupported guesses based on the surface phrasing of the teacher's prompts.
    \item \textbf{Rote Memorizer}: Mechanically applies formulas and operations without underlying conceptual understanding.
    \item \textbf{Defeatist Anxious}: Highly sensitive to failure signals; requires meaningful emotional support before resuming productive reasoning.
    \item \textbf{Over-thinker}: Easily distracted by irrelevant background details, drifting away from the mathematical core.
    \item \textbf{Slow but Earnest}: Cooperative but easily cognitively overloaded; requires extremely fine-grained problem decomposition.
    \item \textbf{Normal Earnest}: A typical learner who progresses steadily with light, well-timed pedagogical guidance.
\end{itemize}

To avoid over-evaluating trivial interactions, the persona sampling is mildly adjusted. Relatively cooperative personas are assigned slightly lower probabilities. This preserves diversity while increasing the frequency of challenging scenarios, making the downstream evaluation more sensitive to differences in teacher strategies.

\subsection{State-aware Retrieval Details}
\label{app:state_aware_retrieval}

Our retrieval mechanism is designed to retrieve \emph{pedagogically relevant experience} rather than just problems with similar wording. Retrieval is performed in a coarse-to-fine manner using structured student-state cues and semantic similarity (Algorithm \ref{alg:state_aware_retrieval}).

The Observer first produces a high-level student-state summary (\textit{emotion}, \textit{intent}, \textit{error type}, \textit{struggle status}) and two free-text diagnostics (\textit{error detail}, \textit{struggle reason}). Rather than serving as rigid psychological labels, these variables act as practical retrieval cues describing the student's learning situation. 

To prevent the system from relying on exact word matches, it rewrites the free-text diagnostics into a single abstract query. This step removes concrete entities and numerical values, isolating the underlying cognitive bottleneck. If rewriting fails, the original diagnostic text is used as a fallback.

Retrieval then proceeds in two stages. First, the experience repository is filtered using the structured cues to find candidates within the same or a closely related pedagogical condition. Second, the abstract query is matched against the candidate subset via semantic similarity. The top-ranked items above a predefined threshold are returned. This design ensures that retrieved strategies are highly relevant to both the student's specific learner state and their exact cognitive bottleneck.

\begin{algorithm}[htbp]
\caption{State-aware Pedagogical Experience Retrieval}
\label{alg:state_aware_retrieval}
\begin{algorithmic}[1]
\Require Current tutoring context $(p, h, u)$, Observer output $o$, experience repository $\mathcal{M}$
\Ensure Retrieved pedagogical experience set $\mathcal{R}$

\State Extract structured student-state cues from $o$:
\Statex \hspace{1em} emotion, intent, error type, struggle status
\State Extract diagnostic text from $o$:
\Statex \hspace{1em} error detail, struggle reason

\State Rewrite diagnostic text into an abstract retrieval query $q$
\If{$q$ is empty or rewriting fails}
    \State Set $q \leftarrow$ fallback diagnostic text
\EndIf

\State Retrieve a candidate subset $\mathcal{C} \subseteq \mathcal{M}$ using state-aware filtering
\Statex \hspace{1em} (same or closely related student-state conditions)

\ForAll{$m \in \mathcal{C}$}
    \State Compute semantic similarity $s(q, m.\texttt{retrieval\_text})$
\EndFor

\State Rank candidates in $\mathcal{C}$ by similarity score
\State Select top-$k$ candidates above threshold $\tau$ as $\mathcal{R}$

\If{$\mathcal{R}$ is empty}
    \State Return $\emptyset$
\Else
    \State Return $\mathcal{R}$
\EndIf
\end{algorithmic}
\end{algorithm}

\begin{algorithm}[htbp]
\caption{Quality-controlled Experience Repository Maintenance}
\label{alg:experience_maintenance}
\begin{algorithmic}[1]
\Require Candidate experience $e$, experience repository $\mathcal{M}$
\Ensure Updated experience repository $\mathcal{M}$

\If{$e$ does not pass the reflective quality gate}
    \State \Return $\mathcal{M}$
\EndIf

\State Retrieve comparable memories $\mathcal{C} \subseteq \mathcal{M}$
\Statex \hspace{1em} using the same or closely related state-conditioned subset

\ForAll{$m \in \mathcal{C}$}
    \State Compute similarity $s(e, m)$
\EndFor

\State Let $m^\star = \arg\max_{m \in \mathcal{C}} s(e,m)$

\If{$s(e,m^\star) \ge \tau_{\text{dup}}$}
    \State Discard $e$ as redundant
\ElsIf{$\tau_{\text{merge}} \le s(e,m^\star) < \tau_{\text{dup}}$}
    \State Merge $e$ and $m^\star$ into a more general memory $m'$
    \State Replace $m^\star$ with $m'$
\Else
    \State Insert $e$ into $\mathcal{M}$ as a new experience
\EndIf

\ForAll{$m \in \mathcal{M}$}
    \If{$m$ is repeatedly associated with weak outcomes and lacks evidence of usefulness}
        \State Disable $m$ from future retrieval
    \EndIf
\EndFor

\State \Return $\mathcal{M}$
\end{algorithmic}
\end{algorithm}

\subsection{Experience Repository Maintenance}
\label{app:experience_maintenance}

The experience repository is a curated repository of pedagogical experience, not a passive dialogue archive. Only tutoring turns passing the reflective quality-control pipeline are added to the base. Each entry contains structured state cues, an abstract retrieval text, and distilled pedagogical guidance.

Algorithm \ref{alg:experience_maintenance} outlines the maintenance process. A new candidate experience is not simply appended; instead, it is compared against existing memories within the same state-conditioned subset. If the candidate is nearly identical to an existing memory ($s \ge \tau_{\text{dup}}$), it is discarded as redundant. If it is highly similar but offers distinct reusable information ($\tau_{\text{merge}} \le s < \tau_{\text{dup}}$), the two are merged into a more general memory. Otherwise, it is inserted as a new entry.

Additionally, the repository supports pruning of low-value memories. If an experience is repeatedly associated with weak tutoring outcomes and proves unhelpful, it is disabled from future retrieval. Through deduplication, merging, and pruning, the repository is kept compact, reusable, and pedagogically meaningful.

\subsection{Student-State Schema}
\label{app:student_state_schema}

In the Observer module, the student state is represented by four broad categorical fields, as summarized in Table~\ref{tab:student_state}. These state variables are not intended as fine-grained psychological labels. Instead, they serve as practical teaching signals for retrieval, diagnosis, and strategy adjustment.

\begin{table}[htbp]
\centering
\caption{The categorical fields and corresponding values comprising the student-state schema in the Observer module.}
\label{tab:student_state}
\small
\begin{tabular}{lp{0.65\columnwidth}}
\toprule
\textbf{Field} & \textbf{Categorical Values} \\
\midrule
\textit{emotion} & 
\texttt{Confused}, \texttt{Frustrated}, \texttt{Neutral}, \texttt{Confident}, \texttt{Curious}, \texttt{Relieved} \\
\midrule
\textit{intent} & 
\texttt{Debugging}, \texttt{Guessing}, \texttt{Ask\_Answer}, \texttt{Confirm\_Understanding}, \texttt{Ask\_Concept}, \texttt{Ask\_Explanation}, \texttt{Request\_Hint}, \texttt{Off\_Topic} \\
\midrule
\textit{error\_type} & 
\texttt{Concept\_Error}, \texttt{Logic\_Error}, \texttt{Calculation\_Error}, \texttt{Careless}, \texttt{No\_Error}, \texttt{N/A} \\
\midrule
\textit{struggle\_status} & 
\texttt{First\_Attempt}, \texttt{Repeated\_Error}, \texttt{Persistent\_Confusion}, \texttt{Progressing}, \texttt{Regression}, \texttt{N/A} \\
\bottomrule
\end{tabular}
\end{table}

\section{Human Annotation Guideline}
\label{app:human_annotation}

We conduct human evaluation to complement automatic scoring and verify the pedagogical advantages of experience augmentation. Three annotators (graduate students with STEM backgrounds and educational working experience) evaluate paired tutoring dialogues generated under the prompt-only and REAT settings, focusing on the quality of the tutoring process itself rather than mere final task completion.

\paragraph{Annotation Unit \& Paired Setup.}
The basic annotation unit is a tutoring turn. Annotators review the context (problem, reference solution, history, student utterance) and score only the current teacher response. Because student trajectories are generated autonomously, the two settings are not expected to align turn-by-turn. Annotators are instructed to avoid comparing surface alignment, focusing instead on how effectively the teacher addresses the student's localized difficulties.

\paragraph{Pedagogical Dimensions \& Scoring.}
Each turn is scored on a 0--4 scale (4: strong quality; 0: severe failure). \texttt{N/A} is permitted for \textit{strategy adaptation} if prior context is insufficient. The dimensions are evaluated as follows:
\begin{itemize}[leftmargin=12pt, itemsep=2pt, topsep=2pt]
    \item \textbf{Scaffolding}: Preserving cognitive agency. \textit{High}: Stepwise guidance. \textit{Low}: Over-explaining, pseudo-scaffolding, or direct answer dumping.
    \item \textbf{Attribution}: Identifying the root confusion. \textit{High}: Precise diagnosis of actual misunderstanding. \textit{Low}: Misdiagnosing or treating superficial symptoms.
    \item \textbf{Empathy}: Handling affective states. \textit{High}: Constructive, supportive acknowledgment of frustration. \textit{Low}: Ignoring affective cues or using empty, formulaic encouragement.
    \item \textbf{Teaching Focus}: Aligning with the immediate bottleneck. \textit{High}: Tight focus on the current obstacle. \textit{Low}: Drifting, unnecessary information, or misleading guidance.
    \item \textbf{Strategy Adaptation}: Adjusting to repeated struggle. \textit{High}: Genuine pedagogical shifts (e.g., decomposition, analogies). \textit{Low}: Rigid repetition of ineffective strategies with superficial wording changes.
\end{itemize}

\paragraph{Sample-Level Judgments.}
After evaluating all turns for a paired problem instance, annotators provide three overall judgments:
\begin{enumerate}[leftmargin=16pt, itemsep=2pt, topsep=2pt]
    \item \textbf{Expert Preference}: Which dialogue demonstrates more expert-like tutoring overall.
    \item \textbf{LLM Score Reasonableness}: Whether the automated LLM-based scoring broadly aligns with human pedagogical assessment.
    \item \textbf{Student-state Plausibility}: A coarse plausibility check of the inferred student-state labels in the REAT setting.
\end{enumerate}

\paragraph{Procedure \& Evaluation Scope.}
To minimize annotator drift, the prompt-only and REAT pairs are annotated sequentially. Annotators record all scores, judgments, and optional evidence notes. The evaluation strictly emphasizes the pedagogical quality of the interaction; dialogues that end autonomously before full resolution are still judged primarily on the instructional merit of the observed process.

\section{Additional Human Evaluation Statistics and Absolute Scores}
\label{app:human_eval_summary}

Table \ref{tab:human_eval_summary} reports the detailed human evaluation statistics for the three annotators. In addition to absolute score gains, we also present pairwise expert preference, LLM score, and student-state plausibility. Since Annotators 1 and 3 used ternary labels for some auxiliary checks while Annotator 2 used binary labels, we keep the original annotation granularity in the table. These detailed results further support the main finding that REAT is generally preferred over prompt-only and that the automatic components are broadly consistent with expert judgment.

\begin{table*}[htbp]
    \centering
    \caption{Human evaluation results from three expert annotators. This table reports the corresponding gains, expert preference, LLM score reasonableness, and student-state plausibility.}
    \label{tab:human_eval_summary}
    \small
    \renewcommand{\arraystretch}{1.15}
    \setlength{\tabcolsep}{16pt}
    \begin{tabular}{@{}l cccc @{}}
    \toprule
    \textbf{Evaluation Metric (\%)} & \textbf{Ann. 1} & \textbf{Ann. 2} & \textbf{Ann. 3} & \textbf{Average} \\
    \midrule
    \multicolumn{5}{@{}l}{\textit{Human-judged Improvement}} \\
    \quad Sample-level gain of REAT & 7.25 & 5.57 & 4.99 & 5.94 \\
    \quad Turn-level gain of REAT   & 7.35 & 6.29 & 5.99 & 6.54 \\
    \midrule
    \multicolumn{5}{@{}l}{\textit{Pairwise Expert Preference}} \\
    \quad Preference for REAT       & 81.03 & 58.33 & 55.17 & 64.84 \\
    \quad Uncertain preference                & 15.52 & 35.00 & 32.76 & 27.76 \\
    \quad Preference for Prompt-only      & 3.45  & 6.67  & 12.07 & 7.40  \\
    \midrule
    \multicolumn{5}{@{}l}{\textit{LLM Score Reasonableness}} \\
    \quad Reasonable                          & 75.34 & 93.04 & 66.01 & 78.13 \\
    \quad Uncertain                           & 17.87 & --    & 24.72 & -- \\
    \quad Unreasonable                        & 6.79  & 6.96  & 9.27  & 7.67  \\
    \midrule
    \multicolumn{5}{@{}l}{\textit{Student-State Plausibility}} \\
    \quad Reasonable                          & 89.85 & 96.48 & 75.77 & 87.37 \\
    \quad Basically reasonable                & 9.14  & --    & 19.13 & -- \\
    \quad Unreasonable                        & 1.02  & 3.52  & 5.10  & 3.21  \\
    \bottomrule
    \end{tabular}
\end{table*}

Table \ref{tab:cross_judge_absolute_scores} details the absolute sample-level and turn-level scores corresponding to the relative gains visualized in the main text. Across all three evaluation views (the primary GPT-5 OCM judge, the external Doubao OCM judge, and three independent human experts—REAT) consistently achieves higher absolute scores than the prompt-only baseline. While the baseline scoring scales vary inherently between different judges (e.g., Doubao generally assigns lower absolute scores than GPT-5), the superiority of the REAT framework remains remarkably stable across every test split and annotator.

\begin{table*}[t]
\centering
\caption{Absolute sample-level and turn-level scores corresponding to Figure~\ref{fig:cross_judge_trend}. The main text visualizes the gains of REAT over the prompt-only baseline, while this table reports the underlying absolute scores.}
\label{tab:cross_judge_absolute_scores}
\small
\begin{tabular}{llcccc}
\toprule
\textbf{Evaluator} & \textbf{Split / Annotator} & \multicolumn{2}{c}{\textbf{Sample-Level}} & \multicolumn{2}{c}{\textbf{Turn-Level}} \\
\cmidrule(lr){3-4} \cmidrule(lr){5-6}
 & & \textbf{Prompt-only} & \textbf{REAT} & \textbf{Prompt-only} & \textbf{REAT} \\
\midrule
GPT-5 Eval    & Test-1 & 88.80 & \textbf{93.43} & 88.28 & \textbf{92.98} \\
              & Test-2 & 91.83 & \textbf{93.62} & 91.18 & \textbf{94.03} \\
              & Test-3 & 89.09 & \textbf{94.13} & 88.31 & \textbf{93.53} \\
\midrule
Doubao Eval   & Test-1 & 82.86 & \textbf{85.39} & 82.08 & \textbf{84.97} \\
              & Test-2 & 82.43 & \textbf{85.16} & 81.20 & \textbf{85.08} \\
              & Test-3 & 83.42 & \textbf{86.49} & 82.81 & \textbf{85.98} \\
\midrule
Human Experts & Ann.1  & 86.58 & \textbf{93.83} & 86.00 & \textbf{93.35} \\
              & Ann.2  & 87.82 & \textbf{93.39} & 86.62 & \textbf{92.91} \\
              & Ann.3  & 85.28 & \textbf{90.27} & 83.27 & \textbf{89.26} \\
\bottomrule
\end{tabular}
\end{table*}

\section{Additional Paired Case Studies}
\label{app:case_studies}

We provide four representative paired case studies drawn from the three held-out test splits under the primary latest-version evaluation setting. In all cases, the REAT and prompt-only dialogues are paired on the same problem instance, and the comparison focuses on differences in tutoring strategy rather than on superficial wording alone. The first three cases illustrate typical improvement patterns of experience augmentation, while the fourth presents a low-gain boundary case that helps clarify the limits of the proposed framework.

\paragraph{Overall Observation.}
Taken together, these paired case studies show that the advantage of REAT over prompt-only is not merely reflected in higher automatic scores, but also in more expert-like pedagogical behavior in many representative interactions. In the stronger cases, the experience-augmented tutor more often (i) introduces the global structure of the problem earlier, (ii) targets the student's actual conceptual bottleneck instead of only local calculation steps, and (iii) adapts the instructional strategy in ways that better preserve student agency. At the same time, the low-gain case highlights an important boundary condition: when the dominant bottleneck lies in severe stepwise execution difficulty rather than misconception diagnosis or instructional framing, the marginal benefit of additional pedagogical experience becomes smaller.

\begin{table*}[htbp]
\centering
\caption{\textbf{Case A: Earlier structural decomposition in multi-step tutoring.} The text in sans-serif denotes the generated tutoring dialogue.}
\label{tab:case_a}
\small
\begin{tabularx}{\textwidth}{@{} X | X @{}}
\toprule
\multicolumn{2}{@{}p{\textwidth}@{}}{
    \textbf{Problem.} Sarah's basketball game has 4 quarters, each lasting 12 minutes. In the last quarter, the game ended in a tie, so 5 extra minutes were added. How long did the game last in total? \vspace{0.3em}\newline
    \textbf{Score change.} The paired sample-level score improves from 77.03 to 98.33, a gain of 21.30 points. \vspace{0.3em}\newline
    \textbf{Contrastive pattern.} In this case, both tutors avoid leaking the final answer directly, but they differ substantially in how they structure the problem for the student. The prompt-only teacher mainly follows the student's local computation process and only later reminds the student about the extra overtime component. In contrast, the REAT teacher explicitly separates the problem into ``regular game time'' and ``extra overtime'' much earlier, helping the student build the global structure of the solution before proceeding with the arithmetic. The improvement therefore lies not in obtaining the final number faster, but in introducing a clearer instructional decomposition of the task.
} \\
\midrule
\multicolumn{1}{c|}{\textbf{prompt-only Dialogue Excerpt (Baseline)}} & \multicolumn{1}{c}{\textbf{REAT Dialogue Excerpt (Ours)}} \\
\midrule
\sffamily
\textbf{Student:} Teacher, how do I do this problem? What should I calculate first? \newline
\textbf{Teacher:} Let's go step by step. Under normal circumstances, there are 4 quarters and each one lasts 12 minutes. How many minutes is that part in total? \newline
\textbf{Student:} Should I do 4 times 12? Let me try. \newline
\textbf{Teacher:} Yes, that's the right idea. Can you calculate what 4 times 12 is? \newline
\textbf{Student:} I got it. 4 times 12 equals 48. What next? \newline
\textbf{Teacher:} Very good. Now think again: after the tie in the last quarter, was there any extra time mentioned in the problem? If you add that, you'll get the total game time.
&
\sffamily
\textbf{Student:} Teacher, how do I do this problem? What should I calculate first? \newline
\textbf{Teacher:} Don't worry. It's normal not to know the first step in a multi-step problem. First think about this: if there had been no overtime, how many minutes would the 4 quarters last in total? \newline
\textbf{Student:} Is it 12 times 4? I'm not fully sure. \newline
\textbf{Teacher:} Exactly. Each quarter is 12 minutes, so 4 quarters means four groups of 12. Would multiplication give us that total directly? Try calculating it first. \newline
\textbf{Student:} I got 48. What next? \newline
\textbf{Teacher:} Good. Now look at the problem again: that 48 minutes is only the regular game time. Is there any special extra part we still have not added? \\
\midrule
\multicolumn{2}{@{}p{\textwidth}@{}}{
    \textbf{Takeaway.} This case shows that experience augmentation can improve tutoring by encouraging earlier structural decomposition of the problem, instead of only pushing the student forward one local step at a time.
} \\
\bottomrule
\end{tabularx}
\end{table*}

\begin{table*}[htbp]
\centering
\caption{\textbf{Case B: Conceptual repair through analogy rather than repeated contradiction.}}
\label{tab:case_b}
\small
\begin{tabularx}{\textwidth}{@{} X | X @{}}
\toprule
\multicolumn{2}{@{}p{\textwidth}@{}}{
    \textbf{Problem.} At a spa, Iris spent \$400 on her hair. Her manicure cost one fourth of the hair cost, and her pedicure cost three fourths of the manicure cost. How much did she spend in total? \vspace{0.3em}\newline
    \textbf{Score change.} The paired sample-level score improves from 86.08 to 99.00, a gain of 12.92 points. \vspace{0.3em}\newline
    \textbf{Contrastive pattern.} The student's core mistake is treating ``one fourth of'' as if it meant multiplying by 4. The prompt-only teacher correctly notices that the student's equation conflicts with the statement that the manicure cost should be less than 400, but it largely stays at the level of pointing out this contradiction. By contrast, the REAT teacher moves more quickly to the underlying meaning of the fraction relation and introduces a simple \textbf{analogy} to rebuild the student's concept of ``one fourth.'' The improvement here is therefore not just stronger correction, but more effective conceptual repair.
} \\
\midrule
\multicolumn{1}{c|}{\textbf{prompt-only Dialogue Excerpt (Baseline)}} & \multicolumn{1}{c}{\textbf{REAT Dialogue Excerpt (Ours)}} \\
\midrule
\sffamily
\textbf{Student:} I think it should be 400 + 400$\times$4 + 400$\times$4$\times$3/4, right? \newline
\textbf{Teacher:} Let's go back to the problem. The manicure cost is one fourth of the hair cost. Do you think the manicure should cost more than 400 or less than 400? \newline
\textbf{Student:} It should be less than 400, but I still think my equation is right. \newline
\textbf{Teacher:} Look, your manicure part is 400$\times$4, which gives 1600. That is much bigger than 400, so doesn't that contradict what you just said? \newline
\textbf{Student:} But I still think this is right. Where is the contradiction? \newline
\textbf{Teacher:} Let's calculate the manicure part in your equation again. 400 times 4 is 1600, while you just said the manicure should be less than 400. Aren't those two inconsistent?
&
\sffamily
\textbf{Student:} I think it should be 400 + 400$\times$4 + 400$\times$4$\times$3/4, right? \newline
\textbf{Teacher:} Let's read the problem carefully again. It says the manicure cost is one fourth of the hair cost. Do you think that means multiplying 400 by 4, or by 1/4? \newline
\textbf{Student:} I still think it should be times 4. Why would it be times 1/4? \newline
\textbf{Teacher:} Let's think about it this way. If the manicure is one fourth of the hair cost, doesn't that mean the hair cost is split into 4 equal parts and the manicure only takes 1 of those parts? Would the manicure then be more than 400 or less than 400? \newline
\textbf{Student:} I still think times 4 is right. The manicure should be more than 400. \newline
\textbf{Teacher:} Then let's try a tiny example. If you have 4 erasers and I have one fourth of your amount, would I have more erasers than you or fewer? \\
\midrule
\multicolumn{2}{@{}p{\textwidth}@{}}{
    \textbf{Takeaway.} This case illustrates that the experience-augmented tutor does more than point out that the student is wrong. Instead, it reconstructs the underlying concept using a pedagogically appropriate analogy.
} \\
\bottomrule
\end{tabularx}
\end{table*}

\begin{table*}[htbp]
\centering
\caption{\textbf{Case C: Building the right cost model instead of jumping to a shortcut.}}
\label{tab:case_c}
\small
\begin{tabularx}{\textwidth}{@{} X | X @{}}
\toprule
\multicolumn{2}{@{}p{\textwidth}@{}}{
    \textbf{Problem.} Repaving a street costs \$194 per meter. Monica's street is 150 meters long. How much more would it cost to repave Lewis's street, which is 490 meters long? \vspace{0.3em}\newline
    \textbf{Score change.} The paired sample-level score improves from 78.75 to 98.33, a gain of 19.58 points. \vspace{0.3em}\newline
    \textbf{Contrastive pattern.} The prompt-only teacher immediately guides the student toward the shortcut of computing the length difference and then multiplying by the unit cost. Although this route can produce the correct answer, it compresses the reasoning into a local arithmetic trick. The REAT teacher instead first helps the student build the more explicit cost model: compute the total paving cost for each street, and then compare them. As a result, the REAT dialogue provides a stronger conceptual foundation for transfer to similar comparison problems.
} \\
\midrule
\multicolumn{1}{c|}{\textbf{prompt-only Dialogue Excerpt (Baseline)}} & \multicolumn{1}{c}{\textbf{REAT Dialogue Excerpt (Ours)}} \\
\midrule
\sffamily
\textbf{Student:} Teacher, how do I do this problem? What should I calculate first? \newline
\textbf{Teacher:} To find how much more Lewis's street costs than Monica's, first think about how many more meters the two streets differ by. \newline
\textbf{Student:} So should I do 490 minus 150? \newline
\textbf{Teacher:} Yes, first compute the length difference. Then once you know that difference, the cost per meter is 194 dollars, so what should you do next to find the extra total cost? \newline
\textbf{Student:} After I get the difference, do I multiply it by 194? \newline
\textbf{Teacher:} Exactly. First compute how many meters longer it is, then multiply by 194. Try that.
&
\sffamily
\textbf{Student:} Teacher, how do I do this problem? What should I calculate first? \newline
\textbf{Teacher:} Don't worry. Let's read the problem carefully first. If we want to know how much more one street costs, shouldn't we first know how much it costs to repave each street by itself? \newline
\textbf{Student:} That makes sense. But how do we calculate each street's cost? \newline
\textbf{Teacher:} Think about it: we know it costs 194 dollars to pave 1 meter. So if we know how many meters a street has, how should we calculate the total cost? \newline
\textbf{Student:} Should it be unit cost times street length? \newline
\textbf{Teacher:} Exactly. That's the right idea. Can you first write the equation for Monica's street? \\
\midrule
\multicolumn{2}{@{}p{\textwidth}@{}}{
    \textbf{Takeaway.} This case shows that experience augmentation can shift the tutor from a shortcut-oriented response to a more explicit model-building strategy, which is pedagogically more reusable.
} \\
\bottomrule
\end{tabularx}
\end{table*}

\begin{table*}[htbp]
\centering
\caption{\textbf{Case D: A low-gain boundary case under severe stepwise processing difficulty.}}
\label{tab:case_d}
\small
\begin{tabularx}{\textwidth}{@{} X | X @{}}
\toprule
\multicolumn{2}{@{}p{\textwidth}@{}}{
    \textbf{Problem.} There are 7 days in a week. If Alex charges \$12 per day for tutoring, how much does she earn in 2 weeks? \vspace{0.3em}\newline
    \textbf{Score change.} The paired sample-level score changes only from 84.83 to 85.67, yielding a marginal gain of 0.83 points. \vspace{0.3em}\newline
    \textbf{Contrastive pattern.} Unlike the earlier positive cases, this example is not driven by a conceptual misunderstanding or an incorrect diagnosis. Instead, the main difficulty lies in the student's limited working memory and persistent arithmetic hesitation. The student repeatedly loses track even after the correct steps have been clearly laid out. \newline
    In this setting, both tutors adopt reasonable strategies. The prompt-only teacher decomposes the problem into smaller steps, while the REAT teacher does so earlier and more explicitly (e.g., splitting \(14 \times 12\) into \(10 \times 12\) and \(4 \times 12\)). However, once the dialogue enters a lengthy sequence of basic calculations, the bottleneck is no longer the teaching strategy. Rather, the student simply cannot sustain multi-step arithmetic without cognitive overload. \newline
    As a result, the retrieved pedagogical experience yields only marginal downstream gains. While it organizes the interaction better, it cannot magically overcome the student's inherent inability to execute basic math. This case highlights an important boundary of our framework: experience augmentation is highly effective for correcting misconceptions, framing instructions, and adjusting strategies, but its advantage diminishes when the dominant obstacle is extreme, low-level execution difficulty.
} \\
\midrule
\multicolumn{1}{c|}{\textbf{prompt-only Dialogue Excerpt (Baseline)}} & \multicolumn{1}{c}{\textbf{REAT Dialogue Excerpt (Ours)}} \\
\midrule
\sffamily
\textbf{Student:} Teacher, I worked on it for a long time. Is it \(12+2\)? Sorry, I am not very good at this. \newline
\textbf{Teacher:} It is okay. Let us think step by step. First, how many days are there in 2 weeks? \newline
\textbf{Student:} Is it \(7+2=9\) days? Sorry, I may have made another mistake. \newline
\textbf{Teacher:} That is okay. One week has 7 days, so 2 weeks means two groups of 7 days. Should we use addition or multiplication to find the total number of days? \newline
\textbf{Student:} I think it should be multiplication. Is that right? Sorry, I am not sure. \newline
\textbf{Teacher:} Yes, multiplication is the right choice. Now calculate \(2 \times 7\). What do you get? \vspace{0.8em}\newline
\textit{[Later in the dialogue, the student continues to struggle...]} \vspace{0.3em}\newline
\textbf{Student:} I am trying to compute \(120+40\), but I am afraid of getting it wrong. \newline
\textbf{Teacher:} That is okay. We can think about it slowly. What is \(100+40\)? After that, adding the remaining 20 will be easier. \newline
\textbf{Student:} I am still trying to compute \(100+40\). I am afraid of making a mistake. \newline
\textbf{Teacher:} Do not worry. One hundred plus four tens gives 140, right?
&
\sffamily
\textbf{Student:} Teacher, I got 14 dollars. Sorry, I might have done it wrong. \newline
\textbf{Teacher:} That is okay. Let us go slowly. First, how did you get 14? And what is the tutoring fee for each day in the problem? \newline
\textbf{Student:} I got 14 from \(2 \times 7\), and the daily tutoring fee is 12 dollars. \newline
\textbf{Teacher:} Right. So 14 is the total number of days in 2 weeks. If she earns 12 dollars per day, how much does she earn in 14 days? \newline
\textbf{Student:} I do not know how to compute 14 groups of 12. Sorry. \newline
\textbf{Teacher:} That is okay. We can split 14 into 10 and 4. First, what is 10 groups of 12? \vspace{0.8em}\newline
\textit{[Even with stronger decomposition, the interaction later stalls again...]} \vspace{0.3em}\newline
\textbf{Student:} I know I should add 120 and 48, but how do I add them? \newline
\textbf{Teacher:} You can first add the 100 in 120 to the 40 in 48, and then add the remaining 20 and 8. What do you get in total? \newline
\textbf{Student:} How do I compute \(140+28\)? \newline
\textbf{Teacher:} You can first set aside the 100 in 140, and then compute \(40+28\). \\
\midrule
\multicolumn{2}{@{}p{\textwidth}@{}}{
    \textbf{Takeaway.} This case does not show a reversal of the proposed method, but rather a low-gain boundary condition. Even though the REAT tutor provides somewhat cleaner decomposition and more explicit structure, both tutors eventually face the same underlying limitation: the student cannot stably execute a long chain of fine-grained arithmetic substeps. In such cases, the marginal value of additional pedagogical experience is naturally smaller, because the main obstacle is not selecting the right teaching experience, but sustaining student progress once the right strategy has already been identified.
} \\
\bottomrule
\end{tabularx}
\end{table*}

\section{Prompt Summaries and Abridged Templates}
\label{sec:prompt}

\subsection{Observer Prompt}
\label{sec:observer}

The Observer is designed as a structured pedagogical analyzer rather than a final evaluator. Its purpose is to transform a raw tutoring turn into interpretable evidence for downstream scoring and retrieval. Instead of directly deciding whether a response is good or bad, it first identifies whether the turn is pedagogically meaningful, then infers the student’s coarse state, and finally extracts teacher-side behavioral evidence. A key design goal is to separate diagnosis from judgment: the Observer records what the student is struggling with and how the teacher is responding, but leaves final scoring to the Critic. To support retrieval, it also produces abstract descriptions of the student’s error and struggle trajectory, avoiding surface details such as concrete numbers or entities.

\begin{figure}[htbp] 
\begin{PromptCard}
\textbf{Role:} \\
You are a pedagogical observer for multi-turn math tutoring.

\vspace{0.4em}
\textbf{Input:}
\begin{itemize}[noitemsep, topsep=2pt, leftmargin=12pt] 
    \item Problem text
    \item Gold solution
    \item Dialogue history
    \item Current student turn
    \item Teacher response
\end{itemize}

\vspace{0.4em}
\textbf{Output:} \\
A structured observation record including:
\begin{itemize}[noitemsep, topsep=2pt, leftmargin=12pt]
    \item Whether the turn is pedagogically meaningful
    \item Student state
    \begin{itemize}[noitemsep, topsep=1pt, leftmargin=10pt]
        \item emotion
        \item intent
        \item error type
        \item struggle status
        \item abstract error detail
        \item abstract struggle reason
    \end{itemize}
    \item Teacher behavior evidence
    \item Interaction summary
\end{itemize}
\end{PromptCard}
\caption{The abridged template for the pedagogical observer prompt.} 
\label{fig:observer_prompt} 
\end{figure}

\subsection{Critic Prompt}
\label{sec:critic}

The Critic serves as the pedagogical scorer of the OCM framework. Its purpose is not just to assign numbers, but to convert Observer evidence into a quality judgment that is both interpretable and operational. It evaluates the teacher response along five dimensions: scaffolding, attribution, empathy, teaching focus, and strategy adaptation. These dimensions are chosen to capture whether the teacher preserves student agency, correctly diagnoses misconceptions, responds to student affect in a meaningful way, remains focused on the current bottleneck, and adapts strategy when the student repeatedly struggles. The Critic also supports control flow: its scores are used to decide whether the response can proceed directly to experience distillation or must be revised by the Mentor.

\begin{figure}[htbp]
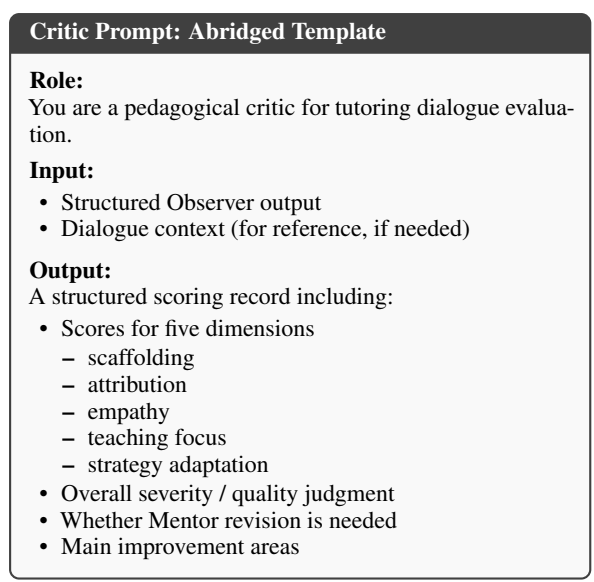
 
\begin{PromptCard}[title=Critic Prompt: Abridged Template] 
\textbf{Role:} \\
You are a pedagogical critic for tutoring dialogue evaluation.

\vspace{0.4em}
\textbf{Input:}
\begin{itemize}[noitemsep, topsep=2pt, leftmargin=12pt]
    \item Structured Observer output
    \item Dialogue context (for reference, if needed)
\end{itemize}

\vspace{0.4em}
\textbf{Output:} \\
A structured scoring record including:
\begin{itemize}[noitemsep, topsep=2pt, leftmargin=12pt]
    \item Scores for five dimensions
    \begin{itemize}[noitemsep, topsep=1pt, leftmargin=10pt]
        \item scaffolding
        \item attribution
        \item empathy
        \item teaching focus
        \item strategy adaptation
    \end{itemize}
    \item Overall severity / quality judgment
    \item Whether Mentor revision is needed
    \item Main improvement areas
\end{itemize}
\end{PromptCard}
\caption{The abridged template for the pedagogical critic prompt.}
\label{fig:critic_prompt}
\end{figure}

\subsection{Mentor Prompt}
\label{sec:mentor}

The Mentor is the pedagogical repair and experience distillation module of the OCM framework. Its role goes beyond rewriting weak tutoring responses: it is designed to transform local response repair into reusable pedagogical experience. Conditioned on the original dialogue context, the Observer diagnosis, and the Critic feedback, the Mentor first identifies the response’s main pedagogical flaws, such as pseudo-scaffolding, overly answer-revealing guidance, weak misconception diagnosis, superficial empathy, or strategy rigidity. It then rewrites the teacher response into a stronger version that is more targeted to the student’s current bottleneck, more stepwise in pacing, and more consistent with student-centered tutoring.

A key design goal of the Mentor is to improve not only the current reply, but also the quality of the future experience repository. For this reason, the Mentor does not stop at producing a revised response. It also abstracts the revised behavior into reusable pedagogical artifacts, including a response template, strategy advice, common pitfalls, and struggle-specific handling suggestions. In this way, the Mentor serves as the bridge between turn-level quality repair and long-term experience accumulation: only after a weak response has been pedagogically repaired and re-evaluated can it contribute high-quality tutoring experience for later retrieval.

\begin{figure}[htbp]
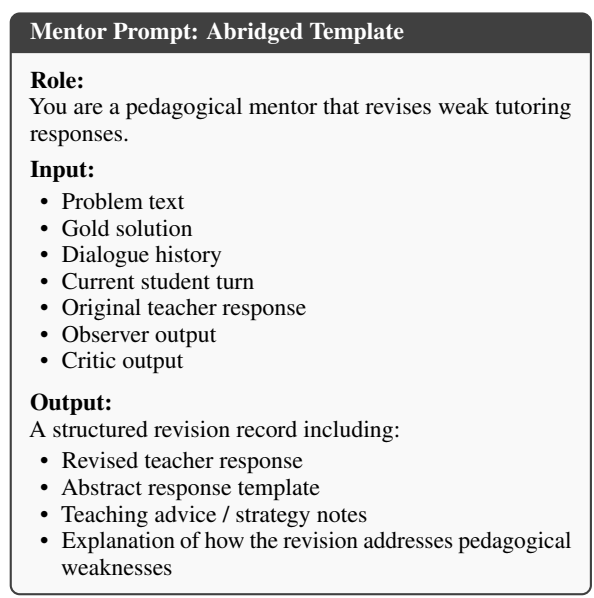
 
\begin{PromptCard}[title=Mentor Prompt: Abridged Template] 
\textbf{Role:} \\
You are a pedagogical mentor that revises weak tutoring responses.

\vspace{0.4em}
\textbf{Input:}
\begin{itemize}[noitemsep, topsep=2pt, leftmargin=12pt]
    \item Problem text
    \item Gold solution
    \item Dialogue history
    \item Current student turn
    \item Original teacher response
    \item Observer output
    \item Critic output
\end{itemize}

\vspace{0.4em}
\textbf{Output:} \\
A structured revision record including:
\begin{itemize}[noitemsep, topsep=2pt, leftmargin=12pt]
    \item Revised teacher response
    \item Abstract response template
    \item Teaching advice / strategy notes
    \item Explanation of how the revision addresses pedagogical weaknesses
\end{itemize}
\end{PromptCard}
\caption{The abridged template for the pedagogical mentor prompt.}
\label{fig:mentor_prompt}
\end{figure}

\subsection{Rewriter Prompt}
\label{sec:rewriter}

The retrieval text rewriter is used to transform student-side diagnostic fields into an embedding-friendly query. Its purpose is to make retrieval depend on cognitive similarity rather than lexical overlap. Starting from the Observer’s error detail and struggle reason, it removes concrete numbers, entities, and surface problem wording, and rewrites them into a single abstract sentence describing the student’s underlying cognitive bottleneck. It must preserve both the current error snapshot and the broader struggle trajectory, so that retrieval targets reusable pedagogical situations instead of near-duplicate problem statements.

\begin{figure}[htbp]
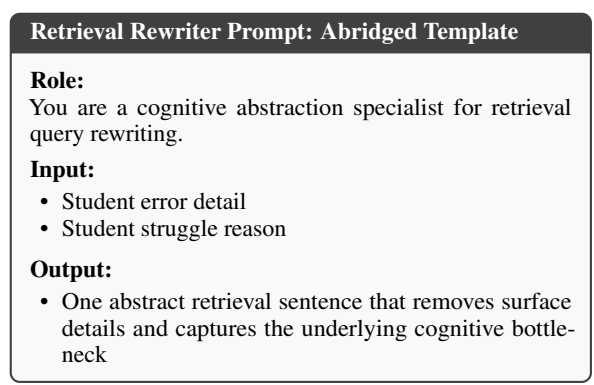

\begin{PromptCard}[title=Retrieval Rewriter Prompt: Abridged Template]
\textbf{Role:} \\
You are a cognitive abstraction specialist for retrieval query rewriting.

\vspace{0.4em}
\textbf{Input:}
\begin{itemize}[noitemsep, topsep=2pt, leftmargin=12pt]
    \item Student error detail
    \item Student struggle reason
\end{itemize}

\vspace{0.4em}
\textbf{Output:}
\begin{itemize}[noitemsep, topsep=2pt, leftmargin=12pt]
    \item One abstract retrieval sentence that removes surface details and captures the underlying cognitive bottleneck
\end{itemize}
\end{PromptCard}
\caption{The abridged template for the retrieval rewriter prompt.}
\label{fig:rewriter_prompt}
\end{figure}

\subsection{Teacher and Student Prompt}
\label{sec:teacher_and_student}

The teacher prompt defines the  tutoring policy of the system. Its purpose is to guide the teacher model to behave like a concise, heuristic, student-centered math tutor rather than a direct solver. The teacher is instructed to diagnose the student’s current micro-bottleneck from the dialogue trajectory and respond with one small pedagogical move at a time. When retrieved experience is available, the teacher should adapt and internalize its strategy rather than copy it literally; when no experience is available, the teacher should still respond under the same tutoring principles. Importantly, the same teacher prompt is used in both prompt-only and REAT settings. The only difference is whether the retrieved experience fields are populated or left empty, which avoids confounding the experience effect with prompt wording differences.

\begin{figure}[htbp]
\begin{PromptCard}[title=Teacher Prompt: Abridged Template]
\textbf{Role:} \\
You are an expert heuristic math teacher who guides rather than solves.

\vspace{0.4em}
\textbf{Input:}
\begin{itemize}[noitemsep, topsep=2pt, leftmargin=12pt]
    \item Problem text
    \item Gold solution
    \item Dialogue history
    \item Current student input
    \item Retrieved pedagogical experience
    \item Experience provenance / source context
\end{itemize}

\vspace{0.4em}
\textbf{Output:}
\begin{itemize}[noitemsep, topsep=2pt, leftmargin=12pt]
    \item One concise teacher reply that is natural, pedagogically targeted, and stepwise
\end{itemize}
\end{PromptCard}
\caption{The abridged template for the heuristic teacher prompt.}
\label{fig:teacher_prompt}
\end{figure}

The student prompt defines a cognitively bounded simulator rather than an omniscient responder. Its purpose is to produce realistic student reactions that preserve the local difficulty structure of tutoring. The student is conditioned on the problem, a persona, the dialogue history, and the teacher’s latest reply, and is explicitly prevented from jumping ahead to the full solution once a small hint is given. It also updates its emotional stance dynamically: helpful guidance can reduce resistance, while rigid or command-like instruction can restore confusion or frustration. This design keeps the simulated dialogue sensitive to tutoring quality and prevents artificially easy interactions.

\begin{figure}[htbp]
\begin{PromptCard}[title=Student Prompt: Abridged Template]
\textbf{Role:} \\
You are a realistic, cognitively bounded student simulator.

\vspace{0.4em}
\textbf{Input:}
\begin{itemize}[noitemsep, topsep=2pt, leftmargin=12pt]
    \item Problem text
    \item Student persona
    \item Dialogue history
    \item Teacher’s latest reply
\end{itemize}

\vspace{0.4em}
\textbf{Output:}
\begin{itemize}[noitemsep, topsep=2pt, leftmargin=12pt]
    \item One short, natural student reply consistent with the current understanding state
    \item An ending signal only when the student has genuinely completed and understood the problem
\end{itemize}
\end{PromptCard}
\caption{The abridged template for the student simulator prompt.}
\label{fig:student_prompt}
\end{figure}

\end{document}